\documentclass[aps,reprint,superscriptaddress,notitlepage]{revtex4-2}
\usepackage[utf8]{inputenc}
\usepackage{amsmath}
\usepackage{physics}
\usepackage[version=4]{mhchem}
\usepackage{url}
\usepackage{amsfonts}
\usepackage{amssymb}
\usepackage{tabularx}
\usepackage{mathtools}
\usepackage{booktabs}
\usepackage{makeidx}
\usepackage{graphicx}
\usepackage{lmodern}
\usepackage{siunitx}
\usepackage{upgreek}
\usepackage{adjustbox}
\usepackage{blindtext}
\usepackage[section]{placeins}
\usepackage{tikz}
\usepackage{tkz-euclide}
\usetikzlibrary{calc}
\usetikzlibrary{optics}
\usepackage{float}
\usepackage{braket}
\usepackage{appendix}
\usepackage{xr}
\begin{document}
\title{Simulation of Axion-Induced Electromagnetic Signal Detection Using Plasmonic Metasurfaces and Diamond NV Centers}
\author{James L. Webb}
\affiliation{Niels Bohr Institute, University of Copenhagen, 2100 Copenhagen, Denmark}%
\email{james.webb@nbi.ku.dk, james.webb42@gmail.com}

\begin{abstract}
The axion represents a strong candidate for weakly interacting dark matter. To date, high sensitivity lab based experiments and astrophysical observations have ruled out a substantial part of the axion mass and photon coupling parameter space. However, a challenge remains in searching for the presence of the axion in the higher mass range 0.01-1eV corresponding approximately to axion field oscillation at THz frequencies. This work investigates via numerical simulation the feasibility of a high sensitivity, lab-based axion sensor operating in this range, based on plasmonic electric field enhancement by a nanostructured metasurface, combined with heterodyne detection and quantum sensing via nitrogen-vacancy (NV) centers in diamond. Estimates of the sensor response to anomalous electromagnetic fields resulting from axion coupling are given using Ti/Au nanopillars on LiNb at axion mass corresponding to telecommunications wavelength ($\approx$0.8eV, 196 THz). Finally, the possibility of sensing in the lower axion mass $<$10$^{-2}$ to 10$^{-1}$eV range is explored using alternative materials, with CdTe as an example.

\end{abstract}
\maketitle

\section{Introduction}

The axion \cite{Wilczek1978} is a promising candidate for a weakly interacting, low mass contributor to dark matter. Arising from a solution to the strong charge-parity problem \cite{Peccei1977,Peccei1977b}, the axion is defined as a pseudoscalar field $a$ weakly interacting with electromagnetic fields according to the Lagrangian density \cite{Wilczek1987}:

\begin{equation}
\mathcal{L}_{a\gamma\gamma} = g_{a\gamma\gamma} a \mathbf{E} \cdot \mathbf{B}
\end{equation}

where $g_{a\gamma\gamma}$ is the axion-photon coupling constant and $a$, $\mathbf{E}$ and $\mathbf{B}$ the axion field, electric field and magnetic field. Through the inverse Primakoff effect, axions can convert to real photons \cite{Sikivie1983}. The consequence of this are modifications to Maxwell's equations \cite{Marsh2016}, producing extremely weak yet potentially detectable anomalous magnetic and electric fields. The modification to the Ampere-Maxwell equation is given by:

\begin{equation}
\nabla \times \mathbf{B} = \mu_0 \mathbf{J} + \frac{1}{c^2} \frac{\partial \mathbf{E}}{\partial t} + \mu_0 c \, g_{a\gamma\gamma} \left( \frac{\partial a}{\partial t} \mathbf{B} + \nabla a \times \mathbf{E} \right)
\end{equation}

The axion field adds an additional effective current density  $\mathbf{J}_{\mathrm{ax}}$. For cold dark matter with velocity only a tiny fraction of the speed of light, the gradient term of $a$ is orders of magnitude smaller than the time derivative $\mathrm{d}a/\mathrm{d}t$. This time derivative term implies an anomalous electric field of magnitude $E_{\mathrm{ax}}$ arising in a strong applied magnetic field of magnitude $B_0$ according to:

\begin{equation}
\mathbf{E}_{ax}(t) \approx -g_{a\gamma\gamma} a(t) \mathbf{B}_0
\end{equation}

with oscillation at the axion Compton frequency $f_a$. The effective current density $\mathbf{J}_{\mathrm{ax}}$ also gives rise to a corresponding time-varying axion magnetic field $\mathbf{B}_{\mathrm{ax}}$.

\begin{figure}[htbp]
    \centering
    \includegraphics[width=0.5\textwidth]{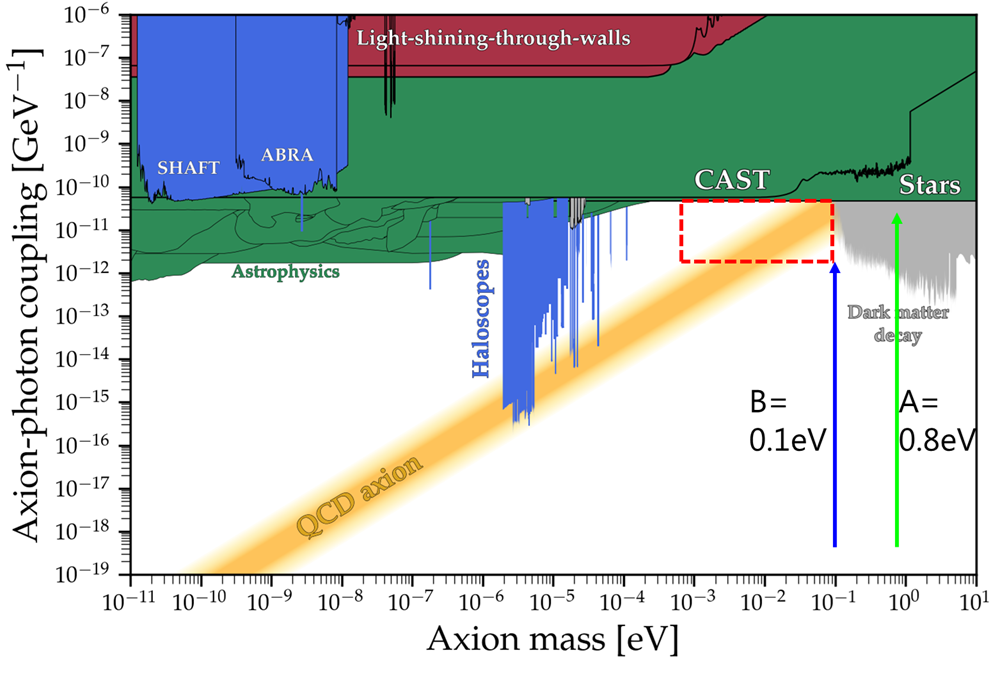}
    \caption{Adapted from \cite{aaaddd}, the regions shaded are those ruled out by previous experiments. The region in the 10$^{-3}$ to 0.1eV range has yet to be extensively explored, yet the theoretical QCD axion prediction range (orange/yellow trace) overlaps with coupling $g_{a\gamma\gamma}$ still relatively high. This work focuses primarily on sensing at label A$\approx$0.8eV (1550nm, 196THz) then applied to the edge of the region at label B$\approx$0.1eV (24THz).}
    \label{fig:figure1}
\end{figure}

Detection of these extremely weak anomalous fields has been the subject of considerable effort over the past decades. Efforts to detect electric field have focused on the haloscope approach such as ADMX \cite{Sikivie1985,Rybka2014,Bartram2021} using resonant shifts in a large integrated volume microwave cavity, with efforts to detect the magnetic field component $\mathbf{B}_{\mathrm{ax}}$, focused on integrating the magnetic flux resulting from the anomalous axion current density $\mathbf{J}_{\mathrm{ax}}$. This has been complemented by alternative experiments such as light shining through walls \cite{VanBibber1987} and the helioscope approach \cite{Schneider1984} in tandem with astrophysical observations \cite{Raffelt2024}. Figure \ref{fig:figure1} shows a portion of the area of the $g_{a\gamma\gamma}$ versus axion mass $m_a$ parameter space ruled out by these past measurements. 

For axion detection, it is necessary to probe a region or volume over which the axion field is coherent, approximately of a scale corresponding to the axion de Broglie wavelength derived from the axion mass and the characteristic velocity of the galactic dark matter halo \cite{Graham2011} . Given the potential viable parameter range for axion mass (10$^{-12}$ to 10$^{7}$eV), this could range from astronomical scale down to the atomic scale. Across the potential mass range, much of the parameter space for photon coupling constants above $g_{a\gamma\gamma}$ =10$^{-10}$ GeV$^{-1}$ has been extensively ruled out through lab-based work and astrophysical observations, which also rule out a large section of the parameter space for higher mass axions ($>$10$^{3}$eV). For reasons of experimental practicality, current lab-based axion detection experiments cluster around the 1-10$\mu$eV mass range, probing a volume on the centimeters to meters scale for a low mass axion field of wavelength many kilometers long. 

There however remains a substantial region ($g_{a\gamma\gamma}$ $<$10$^{10}$, m$_a$$<$10eV) open for detection where standard theoretical axion models KSVZ and DFSZ \cite{Kim1979,Shifman1980} (yellow trace on Figure \ref{fig:figure1}) predict the axion to be found. This higher mass range, between 10meV and 1eV, has been subject to limited study \cite{Roy2023JWSTAxion}, as the practicality of existing lab based methodologies begin to break down in this region. At THz-scale axion field frequency, the signal can become significantly faster than the response rate of many detection systems,  with for example minimal response of atomic or electronic spins to weak axion magnetic fields. Direct THz axion-induced photon detection requires advanced sensors (TES, SNSPDs) rather than well developed SQUIDs and microwave electronics usable at lower axion mass. In the THz range, magnetic susceptibility of many materials drops significantly, making detection of anomalous magnetic fields extremely challenging. Additionally, the coherence scale of the axion field also falls, to the tens of $\mu$m scale at high THz frequencies (eV scale mass), requiring smaller probes of the field to avoid signal cancellation. Scaling down, the haloscope microwave cavity approach suffers multiple problems that render it impractical, including difficulties of building tiny viable microwave cavities, worsening impedance mismatch and amplification bandwidth limitations. Instead, efforts for lab-based axion sensing in the 10meV to 1eV range have focused on alternative approaches such as the use of dielectric stacks \cite{Caldwell2017,Baryakhtar2018}, using parabolic reflectors to focus axion-coupled photons to a detector (e.g. BREAD \cite{Liu2021BREAD}), or coupling the axion field to solid state systems including magnons in antiferromagnetic materials and states in topological insulators \cite{Sekine2021}.  

This work investigates the possibility of building a nanofabricated solid state sensor that can probe this THz, sub-mm scale axion field region, with the goal to sense the anomalous axion contribution to electric field E$_{ax}$. A multistage transduction/conversion process is explored, using 1) a nanofabricated plasmonic metasurface for THz electric field capture and magnification, followed by 2) downconversion to GHz using heterodyne mixing in an electrooptic material with an IR laser, 3) funneling of the resulting signal by a carefully structured coplanar waveguide (CPW) and finally 4) sensing of this signal by shallow NV centers in diamond. The following sections present estimates of the signal strength that can be achieved by this method, making use of finite element simulations to further quantify values and considering primary sources of noise in the system. The integration time required to resolve an axion signal is determined and the limitations of the assumptions made and potential improvements to the scheme are discussed. 

\section{Estimate of Field Strength}

The sensor is positioned in the center of a strong applied magnetic field magnitude $B_0$ aligned along the $x$-direction , maximising the axion electric field component magnitude $E_{ax}$ in the same direction as $B_0$. The field strength is given by the expression:

\begin{equation} 
E_{ax} = \gamma B_0 c \frac{\sqrt{2\rho_a \hbar c}}{m_a c^2 / \hbar} 
\end{equation}

For typical dark matter density estimates, this is approximately \cite{Lee2022TRAX,K2020PhD,green2025doubleresonancestrategyinterferometric,Caldwell2017}: 

\begin{equation}
E_{ax} \approx 10^{-12} \, \text{V/m} \cdot \left( \frac{g_{a\gamma\gamma}}{10^{-15} \, \text{GeV}^{-1}} \right) \cdot \left( \frac{B_0}{1 \, \text{T}} \right) \cdot \left( \frac{100 \, \mu\text{eV}}{m_a} \right)
\end{equation}

This work explores the region between 10meV and 1eV, with $g_{a\gamma\gamma}$ between 10$^{-10}$ and $10^{-12}$ GeV$^{-1}$. This region corresponds to an extremely weak anomalous electric field $E_{ax}$ $\approx$ 10$^{-8}$ to 10$^{-13}$ V/m and axion field coherence length of 1.24mm (for 1eV) up to 12.4cm (for 1meV). In this work, two example frequencies are considered. The first is at $(196~\mathrm{THz},\,1550~\mathrm{nm})$, corresponding to an axion mass $m_{\mathrm{ax}}$ of $\approx$ 0.8eV and a coherence length of 1.6mm. This wavelength is widely used in telecommunications and allows us to leverage knowledge from this field to better define the sensing scheme, including the use of well-known electrooptic effects in lithium niobate (LiNb) for electro-optic effects. The second is 0.1eV ($\approx$ 24THz, 12.4mm coherence length), corresponding to the current edge of astrophysical observation and overlapping theoretical predictions at high $g_{a\gamma\gamma}$. 

\section{Nanoplasmonic Enhancement}

\begin{figure}[htbp]
    \centering
    \includegraphics[width=0.5\textwidth]{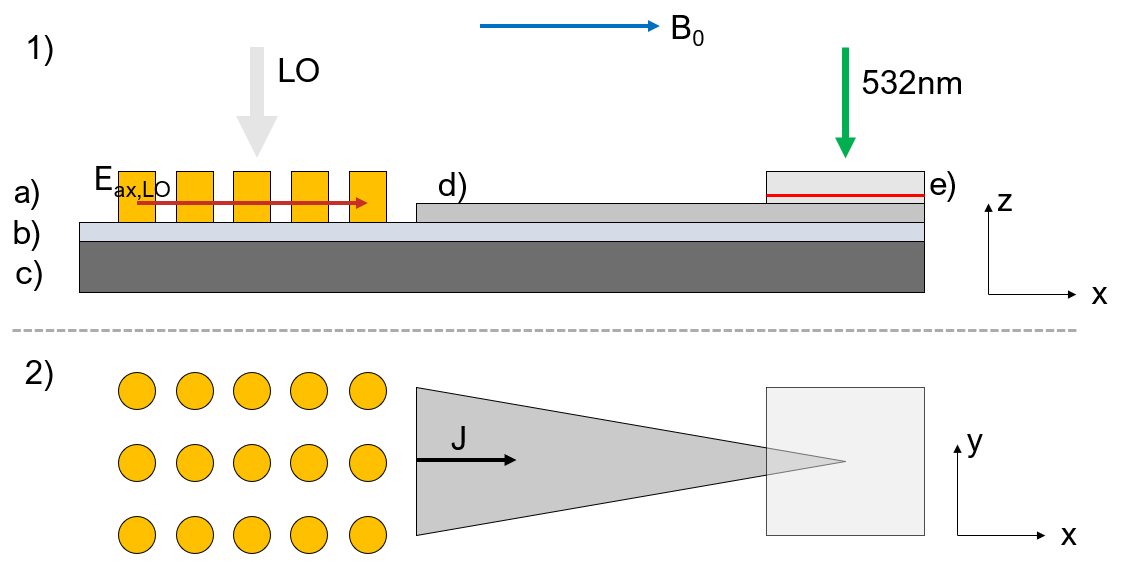}
    \caption{Sketch of the sensor concept (simplified, not to scale) in 1) side-on and 2) top view: an array (a) of nanopillars of size $M \times N$ in the $y$ and $x$ directions, respectively is patterned on thin film electro-optic material b) on a substrate c). Also patterned is a Nb or NbTiN superconducting tapered coplanar waveguide current collector d) onto which a diamond e) containing a thin layer of NV centers (red) is bonded. A strong magnetic field $B_0$ is applied in $x$, with the sensor device in the center of the field. The axion interaction produces an anomalous electric field $E_{\mathrm{ax}}$. which is mixed with the electric field of a polarised infrared local oscillator (LO), injecting a current J proportional to the axion signal strength into a taper, designed to focus the current to a narrow point-like region. The magnetic field induced by this current can then be read out by a layer of NV centres.}
    \label{fig:setup}
\end{figure}

In the first stage, E$_{ax}$ is enhanced using an engineered sub-wavelength metasurface, maximising electric field strength through nanoplasmonic effects in a nanoscale gap between nanofabricated metal pillars on a sapphire substrate layer. Such an approach offers electric field amplitude enhancement \cite{McMahon2011,Ward2010,GarcaMartn2011,Baumberg2019} of factor $E$/$E_0$, with previous studies for surface enhanced Raman scattering showing a realisable enhancement factor in the gap of between 10$^{2}$-10$^{5}$ \cite{Ward2010,Zong2019,Baumberg2019}. Here for simplicity a $M \times N$  square array of Ti/Au metal pillars is modelled, with $M$ nanogaps between each pillar formed in the $y$ direction only and $N$ deep in the $x$ direction  (Figure \ref{fig:setup}). Spatial phase coherence is assumed to be maintained across our nanopillar array, which in turn constrains the maximum array, pillar and gap size to the coherence scale of the axion field. Such an array can be readily nanofabricated, using electron beam lithography or nanoimprint lithography. The surface is aligned parallel to applied field $B_0$ such that the direction of $E_{\mathrm{ax}}$ aligns with the $x$-direction, parallel to the plane of the substrate and aligned with the direction of the nanogaps. The pillars are designed to enhance the field with a peak at the particular target detection frequency corresponding to the axion mass. For other frequencies, pillar design can be simulated and changed in the nanofabrication process.  

To calculate the degree of enhancement for a single gap, FDTD simulations (Ansys High Frequency Simulation Software) are used to obtain $E_{\mathrm{LO}}$ and $E_{\mathrm{ax}}$ and determine the gap enhancement factor $E$/$E_0$ as a function of pillar gap, radius and height at 196THz.  The pillar dimensions at peak enhancement, radius $r$=60nm and height $h$=280nm correspond to those typical in the literature, with enhancement factor $E$/$E_0$=480.4. The factor can be estimated using the phenomenological relation: 

\begin{equation}
\frac{E_{ax}}{E_0} = Q \left(\frac{r}{g}\right)^{\alpha} \left(\frac{h}{r}\right)^{\beta}
\end{equation}

with $\alpha$, $\beta$= 0.5-1 and the Drude plasmon quality factor \cite{Maier2007Plasmonics} Q =-Re($\epsilon_m$)/Im($\epsilon_m$), taking $\epsilon_m$=-115+11i for gold at 196THz. 

\begin{figure}[htbp]
    \centering
    \includegraphics[width=0.5\textwidth]{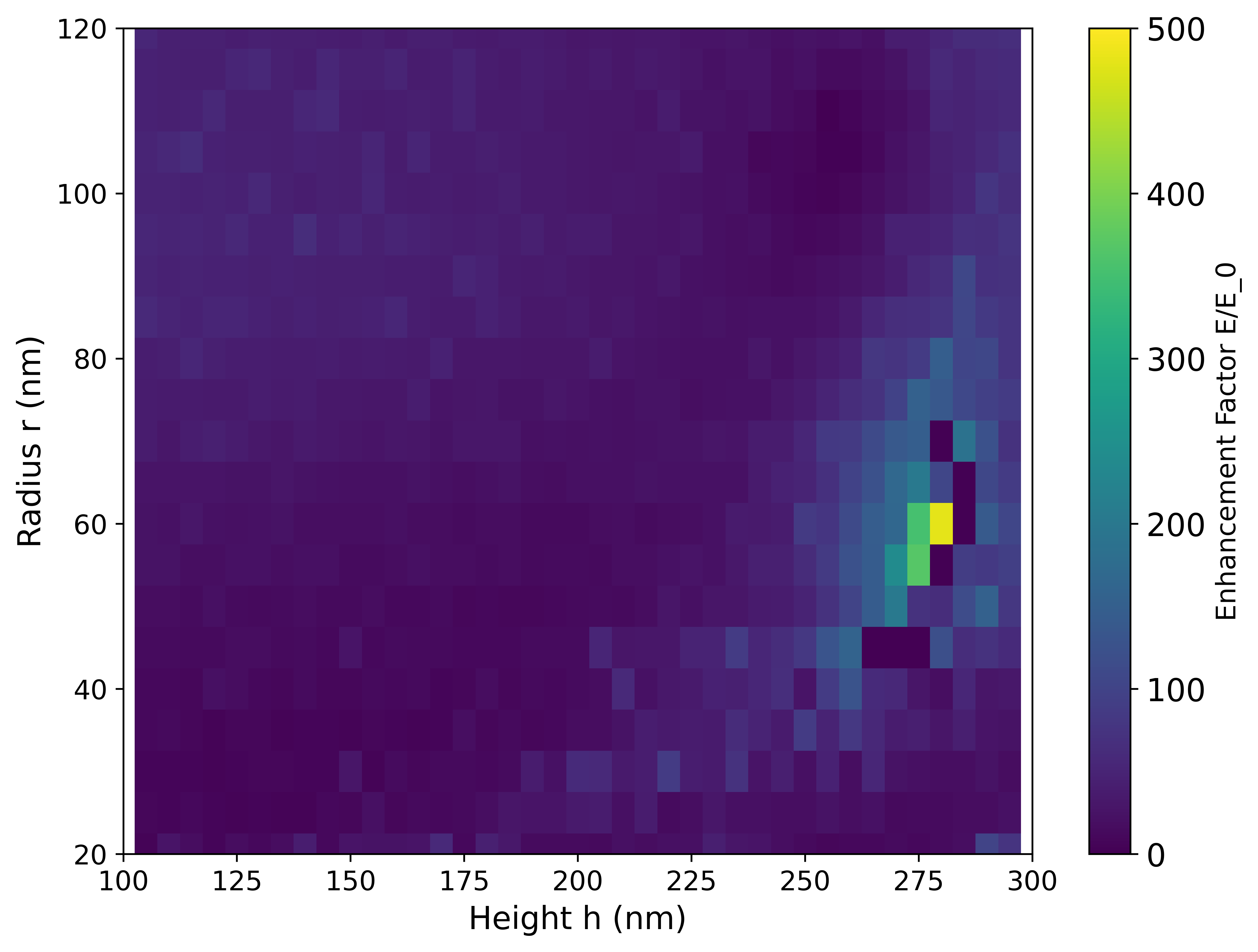}
    \caption{Enhancement factor $E$/$E_0$ for Ti/Au pillars on LiNb at 196THz, varying pillar radius and height. Maximum enhancement factor 480.4 is reached at radius $r$=60nm, height $h$=280nm.}
    \label{fig:pmap}
\end{figure}

\begin{figure}[htbp]
    \centering
    \includegraphics[width=0.48\textwidth]{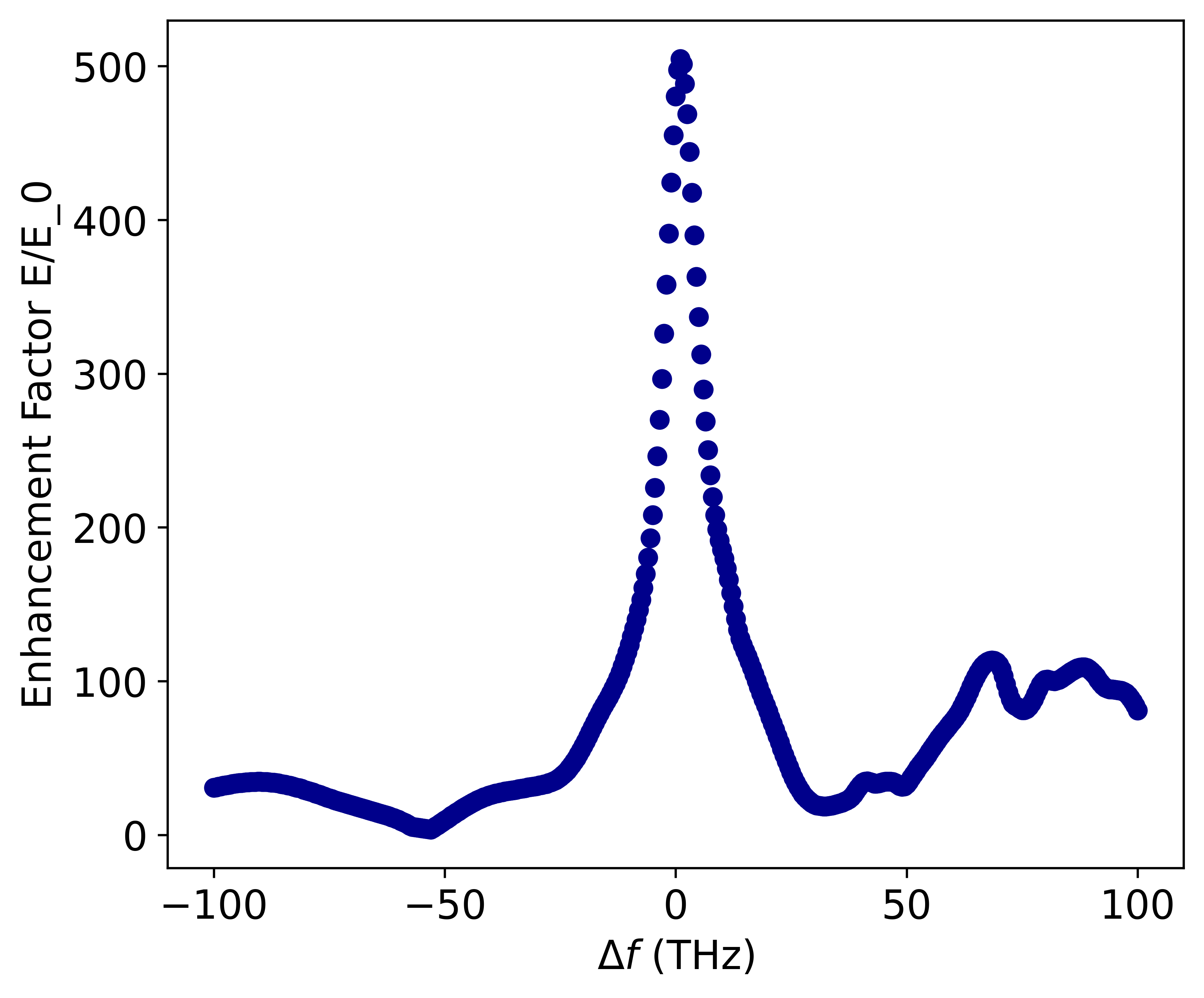}
    \caption{Simulation of the enhancement factor E/E$_0$ in a nanogap as a function of axion field frequency deviation from 196THz. The peak has a FWHM of 7THz and Q=28.6.}
    \label{fig:fsweep}
\end{figure}

In Figure \ref{fig:fsweep} the response at radius and height for peak enhancement as a function of frequency $\Delta f$ away from 196 THz is shown, to explore the degree of tunability available in the scheme. The metasurface effectively acts as a bandpass filter to the axion signal. Enhancement drops by a factor of 5 within $\Delta f$=10THz. The surface structure along with the local oscillation wavelength tunability defines the detection frequency range of the sensor, with alternative dimensions required for targeted detection of different axion mass.  

\section{Heterodyne with Local Oscillator}
%section checked 27/6, seems valid

In order to amplify and downconvert the signal into the useable GHz range, this work models using a laser as a local oscillator (LO) source of electric field strength $E_{\mathrm{LO}}$ to perform heterodyne difference frequency generation (DFG) with the oscillating axion signal ($E_{\mathrm{ax}}$) to an intermediate frequency (IF) $f_{\mathrm{IF}}$=30.89GHz that corresponds to the $m_s$=0 to $m_s$=1 NV center transition. $f_{\mathrm{IF}}$ is given by:

\begin{equation}
f_{\text{IF}} = |f_{\text{sig}} - f_{\text{LO}}|
\end{equation}

\begin{equation}
I_{\text{IF}}(t) \propto \cos(2\pi f_{\text{IF}} t + \phi_a(t) - \phi_{\text{LO}})
\end{equation}

Conversion between 20-30GHz and THz signals has previously been explored in reverse for radio-over-fiber applications, converting from 5G/6G wifi to telecoms wavelengths using LiNb electro-optic modulation \cite{Martin2017,10521695} It should be noted that $E_{\mathrm{LO}}$ and $E_{\mathrm{ax}}$ will not be phase coherent, as the coherence time of $E_{\mathrm{ax}}$ at THz is only 4ns and neither are phase locked. Readout therefore relies on measuring signal power, discarding phase information and relying on scaling as the square root of the overall measurement time $\sqrt{T_{\mathrm{meas}}}$ The laser of power $P_{\mathrm{LO}}$ is assumed to be distributed to cover each nanogap with even intensity across the array, resulting in the same local oscillator field $E_{\mathrm{LO}}$ at each gap. The laser is linear polarised to align with $E_{\mathrm{ax}}$ in the gap. The laser is assumed to be aligned such that there is a closely similar degree of enhancement in the nanogap region for both $E_{\mathrm{ax}}$ and $E_{\mathrm{LO}}$, with the strongest mixing in the LiNb layer between the pillars. Finally, pulsing the laser is considered, to realise as high a peak electric field as possible, with pulse length $\tau_{\mathrm{pulse}}$ and repetition rate $f_{\mathrm{rep}}$.

\section{Downconversion Layer}

For DFG to $f_{\mathrm{IF}}$=30.89GHz, electro-optic effects in standard optoelectronic materials can be used. In such materials, the polarisation $P$ generated by an electric field is given by \cite{Yoshioka:24,boyd2020nonlinear}:

\begin{equation}
P = \epsilon_0 \left( \chi^{(1)}E + \chi^{(2)}E^2 + \chi^{(3)}E^3 + \dots \right)
\end{equation}

where $E$ is electric field and $\chi^{1}$ and $\chi^{2}$ are the first and second order components of the electric susceptibility. For second order processes commonly exploited for mixing, polarisation is given by:

\begin{equation} 
\mathbf{P}^{(2)}_{\text{total}} = -\int_{V} \epsilon_0 \, n^4(\mathbf{r}) \, r_{\text{eff}}(\mathbf{r}) \, E_{\text{ax}}(\mathbf{r}) E_{\text{LO}}(\mathbf{r}) \, dV 
\end{equation}

where $r_{\mathrm{eff}}$ is the effective linear electro-optic coefficient. The current of the $f_{\mathrm{IF}}$=30.89GHz IF signal produced is linearly proportional to the amplitude of the polarisation and the IF frequency, as $J_{\mathrm{IF}} = \frac{\mathrm{d}P}{\mathrm{d}t}$:

\begin{equation}
\mathbf{J}_{IF}(\mathbf{r}) = \frac{\partial \mathbf{P}^{(2)}(\mathbf{r})}{\partial t} = -i \omega_{IF} \mathbf{P}^{(2)}(\mathbf{r})
\end{equation}

Localising the field in the nanogaps as RF sources gives the advantage of achieving good localised phase matching, allowing a strong degree of constructive contribution to the IF signal. The disadvantage of this method is that although $r_{\mathrm{eff}}$ is large in the context of electro-optic effects, it is extremely small for converting the tiny axion signal. For LiNb in an X-cut or Y-cut configuration to align maximum response to the nanogaps, $r_{\mathrm{eff}}$=20-30pm/V. This considerably increases the required LO power. An alternative is to use materials with stronger effects, for example barium titanate (BTO) has $r_{\mathrm{eff}}$=300-900pm/V \cite{abel2019large}. However, many such materials suffer far higher dielectric loss, with dielectric loss tangent in LiNb $\tan \delta$ $\approx$10$^-3$ at T=300K and $\approx$10$^-4$ at 77K \cite{zhou1995microwave} compared to BTO with $\tan \delta$ = 0.01-0.05 \cite{tagantsev2003intrinsic}. This has significant impact on the noise budget of the scheme (see later section). A further alternative is to use transition metal dichalcogenide (TMD) monolayers such as MoS$_2$, WS$_2$ or WSe$_2$ where $r_{\mathrm{eff}}$ can be up to 10$^4$, but with the major disadvantage of greatly reduced signal generation volume\cite{li2013probing,janisch2014extraordinary}.

By the DFG process, $E_{\mathrm{ax}}$ generates 30.89GHz sidebands when mixed with the laser field $E_{\mathrm{LO}}$ along the correctly cut principal crystal axes. The resulting polarisation where the electric fields are the peak field in V/m can be modelled as:

\begin{equation}
P^{(2)}(IF) = \varepsilon_0 \chi^{(2)}_{\mathrm{eff}} E^{*}_{LO} E_{ax}
\end{equation}

with: 

\begin{equation}
\chi^{(2)}_{\mathrm{eff}} = -n^4 r_{eff}
\end{equation}

where the effective $\chi_{\mathrm{eff}}^2$ is derived from the refractive index $n$ and the relevant nonlinear coefficient $r_{\mathrm{eff}}$. Using the above equations local polarisation and resulting induced nonlinear current density $J_{\mathrm{IF}}$ can be calculated for a given factor of enhancement of electric field, scaling in from the simulation to the real values of $E_{\mathrm{ax}}$ and $E_{\mathrm{LO}}$ respectively. 

The advantage of the DFG approach is so that the LiNb can simply be used as a highly localised nonlinear mixer in tandem with the metasurface enhancement. Alternative schemes that propose using electro-optic effects in LiNb directly for optical detection such as in \cite{green2025doubleresonancestrategyinterferometric}, GALILEO and CARAMEL \cite{Ebadi_2024,Davoudiasl2026} face issues with the strong dispersion using long path lengths in LiNb, needing carefully designed and increasingly small cavities and periodic poling that places upper mass limits in the $\approx$10$^{-4}$eV range. In addition, THz detection requires stable frequency reference comparison and specialized detectors working in the THz regime. DFG allows the axion signal to be addressed in the GHz microwave domain, with well-developed electronics and sensing. 

\section{Current Collection}

The IF current created by DFG across the array must be effectively captured with as low noise as possible. A single nanogap acts as a high capacitance current source with high overall impedance $Z_{\mathrm{gap}}$, high reactance $X_{\mathrm{gap}}$ associated with the capacitance of each nanogap, and also a high resistance $R_{\mathrm{gap}}$ governed by the electro-optic layer (LiNb) dielectric loss. For a single gap, this source poses some difficulty with a typical low impedance microwave waveguide, as the majority of the IF signal would be lost due to impedance mismatch and with a high Johnson-Nyquist noise. The gap is defined as a parallel RC circuit with impedance as the inverse of the admittance $Z_{\mathrm{gap}} = \frac{1}{Y_{\mathrm{gap}}}$, where:

\begin{equation}
Y_{gap}=\frac{1}{R_{gap}}+j \omega C_{gap}.
\end{equation}

Gap resistance $R_{\mathrm{gap}}$ is defined in terms of the dielectric loss tangent tan $\delta$ as:

\begin{equation}
R_{gap}=\frac{tan \delta}{2\pi f_{IF}}
\end{equation}

Substituting expressions, for typically small $\tan \delta$ = 10$^{-3}$ to 10$^{-5}$, the real and imaginary parts of the gap impedance can be given as:

\begin{equation}
Re(Z_{gap})=\frac{tan \delta}{2\pi f_{IF}C_{gap} (1+ tan^{2} \delta)} \approx \frac{tan \delta}{2\pi f_{IF}C_{gap}}
\end{equation}

\begin{equation}
Im(Z_{gap})=-\frac{1}{2\pi f_{IF}}
\end{equation}

$Z_{\mathrm{gap}}$ can be modelled analytically an approximate value for the gap capacitance using the facing area of each pillar $A_{\mathrm{face}}$ and a finite penetration depth into the LiNb. $C_{\mathrm{gap}}$ is defined as the sum of the fraction of the pillar height that is air gap plus the fraction in the LiNb, each given by: 

\begin{equation}
C_i = \frac{\pi \epsilon_0 \epsilon_{r,i} h}{\operatorname{arccosh}\left(1 + \frac{g}{2r}\right)}
\end{equation}

with i=air or i=LiNb and $\epsilon_{r_i}$ the relative permittivity of air and LiNb. Gap resistance can also be modelled  analytically as:

\begin{equation}
R_{\text{gap}} = \frac{1}{\omega C_{gap} \tan\delta}
\end{equation}

$Z_{\mathrm{gap}}$ from this analytical form can be compared with the value determined by FDTD simulation in Ansys.

Although a single gap is highly mismatched, this can change as the array is scaled up. In scaling $M \times N$, we make several assumptions for the behaviour of the array. First, it is assumed that all nanogaps are identical and the array can be coupled into a current capture structure (e.g. a CPW) such that it sees a uniform waveguide electric field across it, with minimal fringing or edge effects. It is also assumed that all nanogaps produce a coherent RF signal (valid when array size $<$ axion coherence length and array size $<<$ IF wavelength). In this limit, the array can be treated as a single admittance sheet producing a current $J_{\mathrm{IF}}$ in a single direction defined by the gap orientation and laser polarisation. In these limits, $J_{\mathrm{arr}} \propto J_{\mathrm{gap}} \times (M \times N)$ with each additional added nanogap adding current coherently and with total capacitance $C_{\mathrm{arr}} \propto C_{\mathrm{gap}} \times (M \times N)$ also increasing. This capacitance is assumed to be significantly larger than parasitic capacitance. At the same time, source resistance as seen by a collector $R_{\mathrm{arr}} = \frac{R_{\mathrm{gap}}}{M \times N}$ is reduced by adding more nanogaps. As $M \times N$ scales upwards, it is assumed the total array current can be written as:

\begin{equation}
I_{arr} = I_{gap} M N \eta_{c} (\tau_{pulse}  f_{rep})
\end{equation}

where $\eta_{\mathrm{c}}$ is the structural fill factor, $\tau_{\mathrm{pulse}}$ is the LO laser pulse length, and $f_{\mathrm{rep}}$ is the laser repetition rate.

The structural fill factor $\eta_{\mathrm{c}}$  arises from the difference in scale between the nanogaps where the IF signal field is generated and the much larger size of the structure that captures it (mode overlap between the gap and structure field). The precise value can be determined by integrating the mode overlap of the gap sources and the waveguide current and electric field via a complete field and electro-optic simulation of both. Here the structural fill factor represents a simplification of this for the assumptions taken (e.g. uniform waveguide E-field, uniform gap emitters). For a CPW this can be approximated by the ratio of the width $\eta_{\mathrm{c}}$  $\approx$ M$\times$w$_g$/W$_{CPW}$. For a CPW of 125$\mu$m, M=1000 and w$_g$=10nm, this gives $\eta_{\mathrm{c}}$ $\approx$ 0.08. 

It can be noted that under these assumptions that coupling into a current capturing structure in the normal state becomes unfeasible as the thermal noise from any waveguide structure would dominate the total noise from the array. Both this and the high admittance implies that a superconducting structure is required to capture the array current. This can be achieved by a planar structure made from Nb or NbTiN carefully oriented in the ($x$,$y$) plane, such that the flux from the applied field $B_0$ does not significantly penetrate the thin film superconducting layer. 

In this work the simplest collection structure is modelled, consisting of a superconducting coplanar taper, of starting width $w_s$=200$\mu$m, end width $w_e$=50nm and start/end gap width $s_{\mathrm{gap}}$ and $e_{\mathrm{gap}}$ with the array modelled to be directly at the taper wide end mouth. As it is not computationally feasible for us to simulate both the macroscopic current capture structure and many hundreds of nanogaps in a single finite element model, current capture is simulated by approximating the array as a lumped port impedance $Z_{\mathrm{arr}}$ at the taper mouth, and calculating $I_{\mathrm{start}} / I_{\mathrm{end}}$ as taper geometry is varied, also taking into account the kinetic inductance of the Nb superconductor layer. 

\begin{figure}[htbp]
    \centering
    \includegraphics[width=0.5\textwidth]{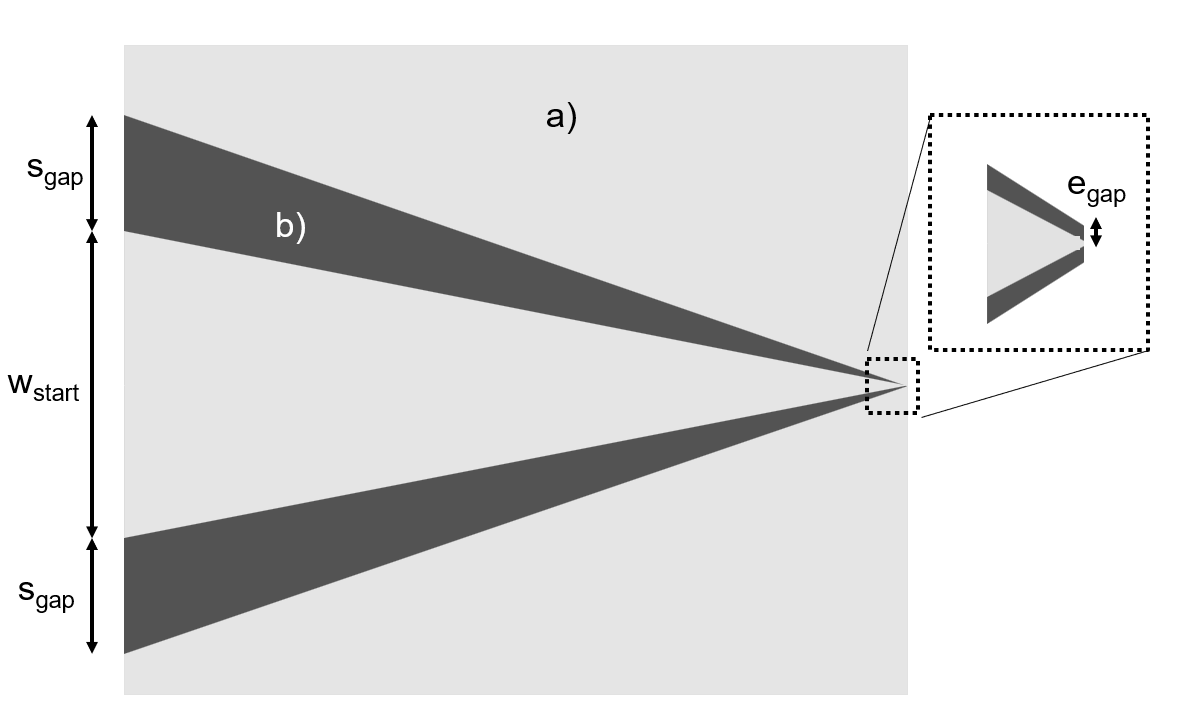}
    \caption{Sketch (not to scale) of relevant taper dimensions, including start central conductor width $w_s$, CPW gap $s_{gap}$, taper end width $w_e$ and taper end gap $e_{gap}$. }
    \label{fig:tapergeo}
\end{figure}

\begin{figure}[htbp]
    \centering
    \includegraphics[width=0.5\textwidth]{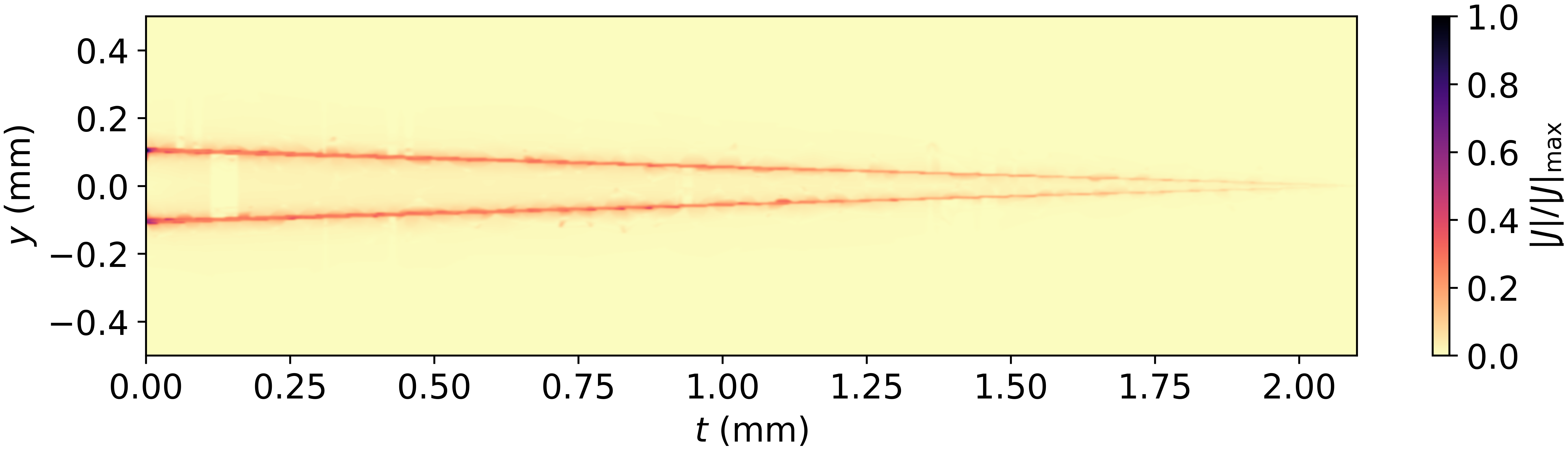}
    \caption{Taper length $t$=2.1mm, corresponding to the best impedance match between the array and taper. This results in the highest input current $I_{\mathrm{start}}$ and current at the taper end $I_{\mathrm{end}}$, but not the optimal transfer $I_{\mathrm{start}} / I_{\mathrm{end}}$. Current density is highest towards the edges of the central conductor, as expected for a superconducting CPW \cite{Marchiori2022,Beck2016}. }
    \label{fig:taperplot_a}
\end{figure}

\begin{figure}[htbp]
    \centering
    \includegraphics[width=0.5\textwidth]{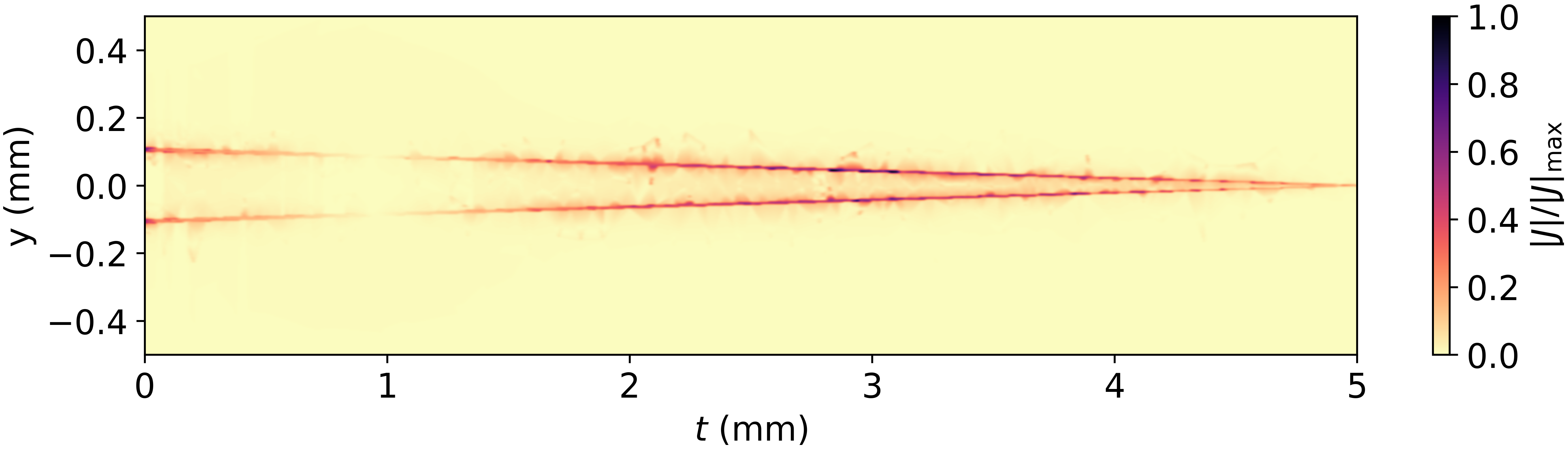}
    \caption{Taper length $t$=5mm, corresponding to the best ratio $I_{\mathrm{start}} / I_{\mathrm{end}}$, with a current antinode formed at the taper end. }
    \label{fig:taperplot_b}
\end{figure}

\begin{figure}[htbp]
    \centering
    \includegraphics[width=0.5\textwidth]{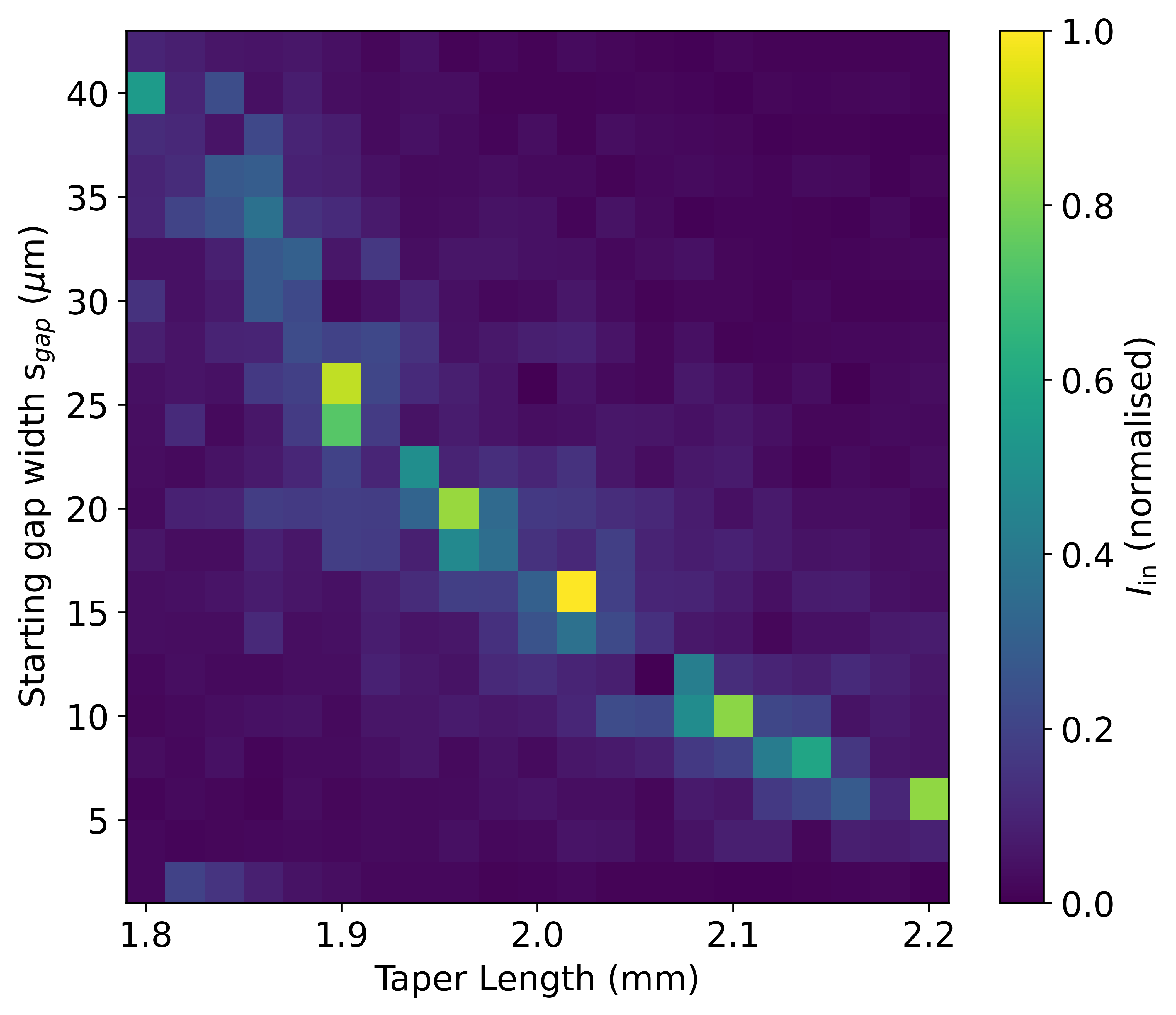}
    \caption{Normalised peak input current density $J_{\mathrm{start}}$ at the start of the taper as a function of the starting taper to ground plane gap $s_{\mathrm{gap}}$ ($\mu$m) and taper length $t$ (mm) as calculated from Ansys HFSS simulation. The gap at the end of the taper is fixed at $e_{\mathrm{gap}}$=2$\mu$m. The peak current input arises when the geometry makes the reactive component of taper input impedance matched to the array i.e. Im($Z_{\mathrm{arr}}$)=-Im($Z_{\mathrm{taper}}$). }
    \label{fig:taper1}
\end{figure}

\begin{figure}[htbp]
    \centering
    \includegraphics[width=0.48\textwidth]{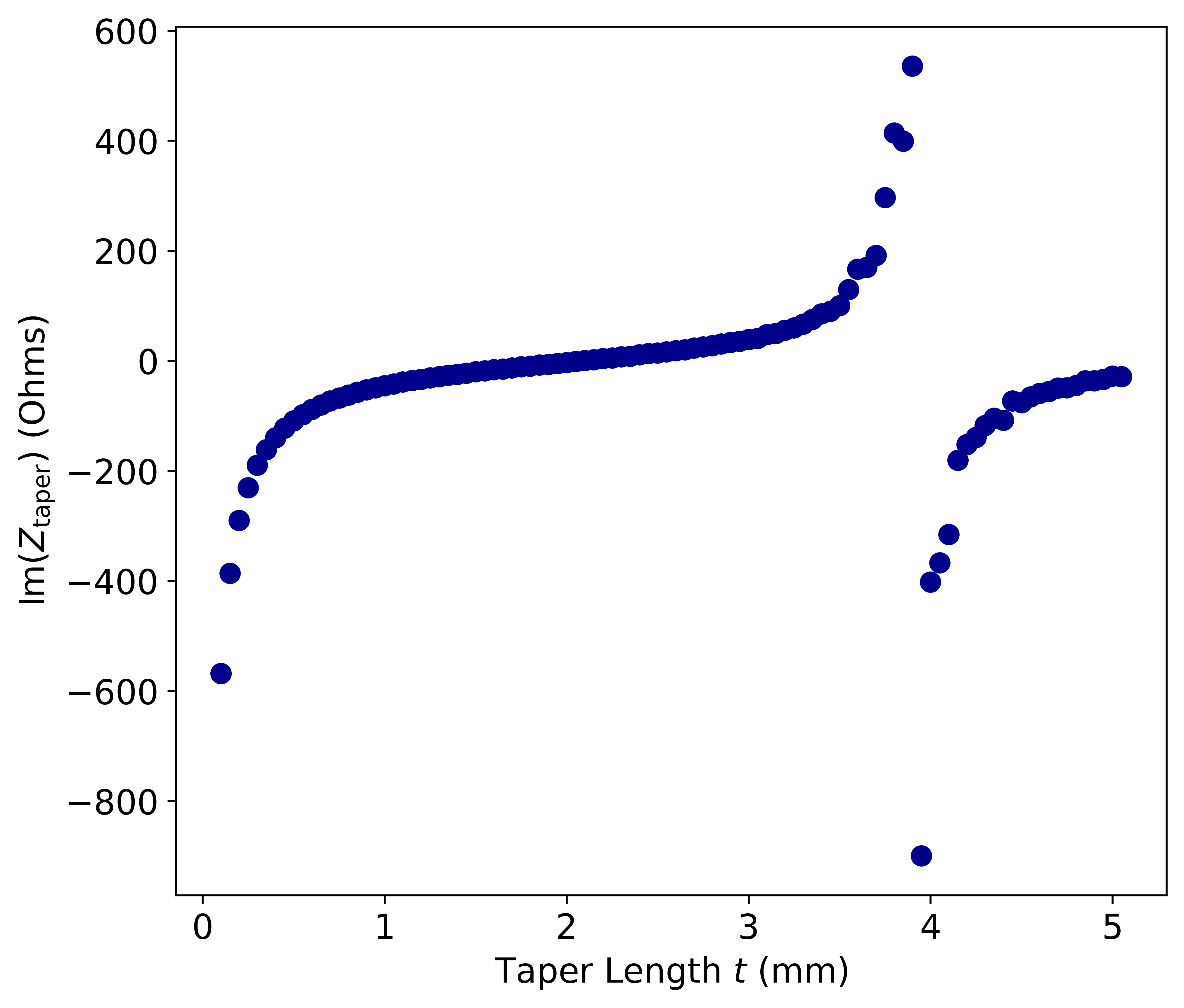}
    \caption{The imaginary component of current capture coplanar taper input impedance $Z_{\mathrm{taper}}$ as calculated from Ansys HFSS simulation plotted against taper length $t$(mm) for $s_{\mathrm{gap}}$=12$\mu$m, $e_{\mathrm{gap}}$=2$\mu$m }
    \label{fig:taper2}
\end{figure}

\begin{figure}[htbp]
    \centering
    \includegraphics[width=0.48\textwidth]{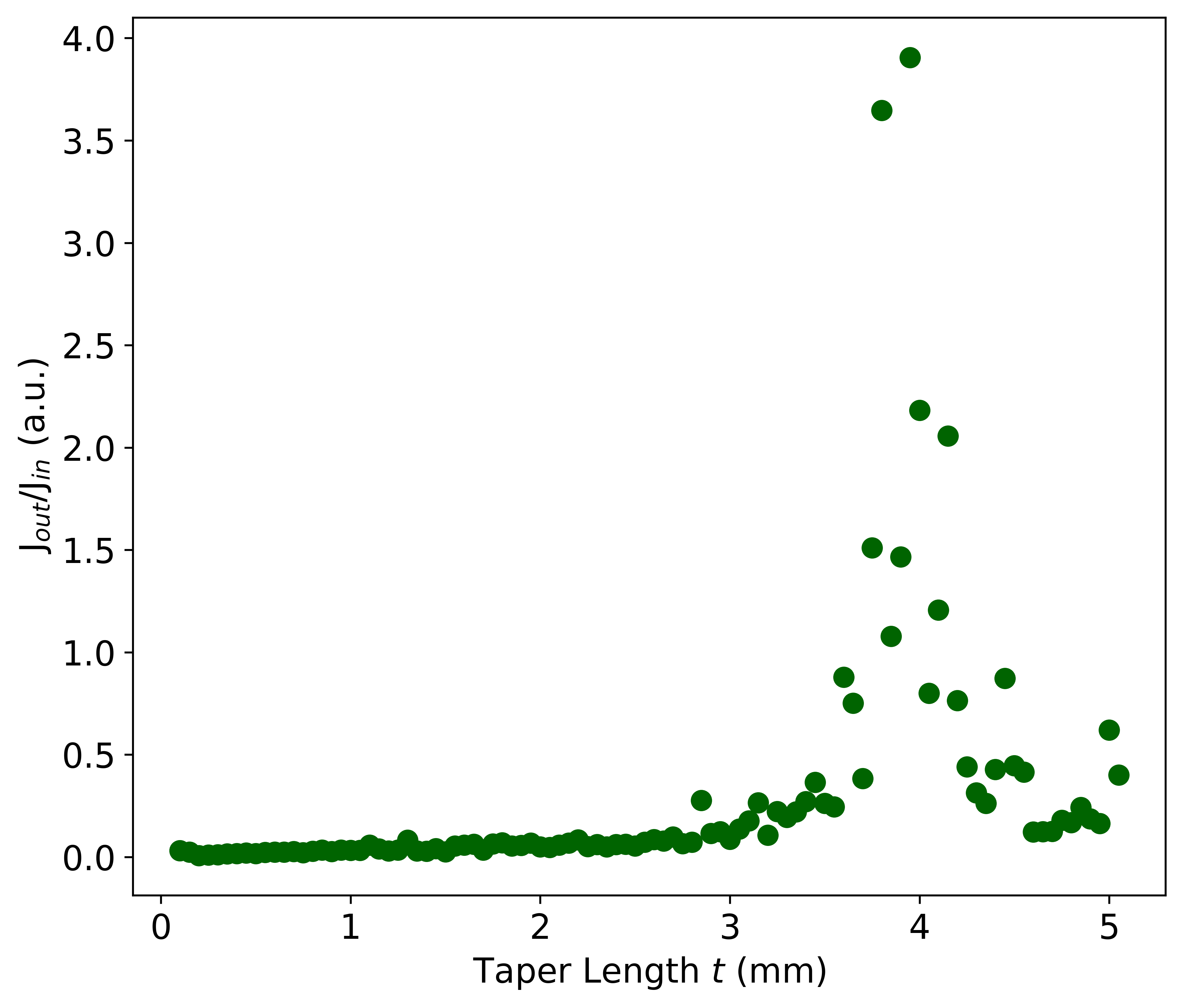}
    \caption{The ratio of current density at the start of the taper $J_{\mathrm{start}}$ to that at the end of the taper $J_{\mathrm{end}}$ as a function of taper length $t$(mm) for $s_{\mathrm{gap}}$=12$\mu$m, $e_{\mathrm{gap}}$=2$\mu$m. }
    \label{fig:taper3}
\end{figure}

\begin{figure}[htbp]
    \centering
    \includegraphics[width=0.48\textwidth]{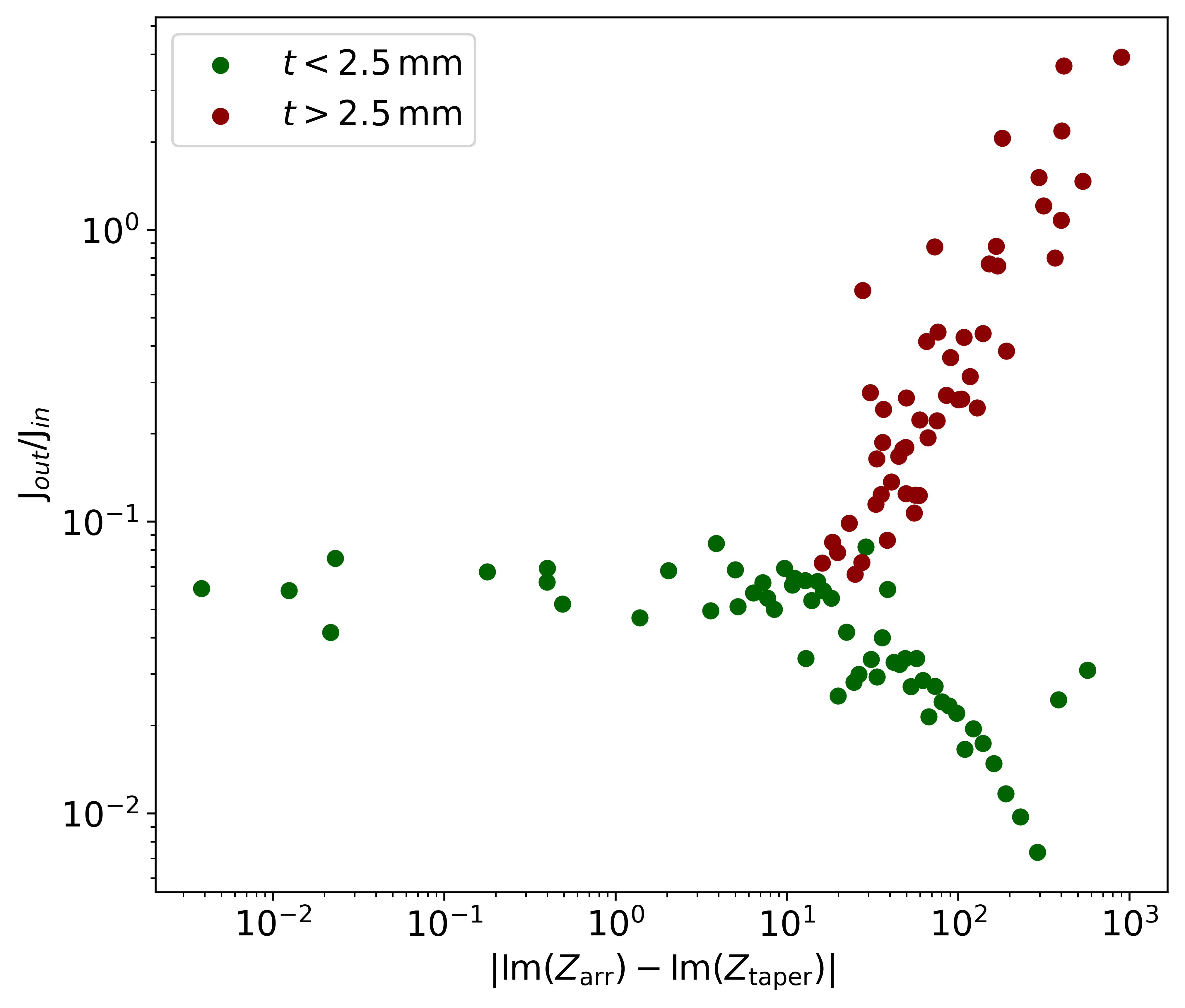}
    \caption{The current ratio  $J_{\mathrm{start}} / J_{\mathrm{end}}$ plotted against absolute reactance difference $Z_{\mathrm{arr}}$-$Z_{\mathrm{taper}}$ for $s_{\mathrm{gap}}$=12$\mu$m, $e_{\mathrm{gap}}$=2$\mu$m. Plotted separately is the data for taper length $t<$2.5mm and $t>$2.5mm.}
    \label{fig:taper4}
\end{figure}

We model the taper in Ansys as a Nb coplanar waveguide, with a central conductor tapered in the $x$-direction of starting width $w_{s}$=200$\mu$m, starting gap width to the ground plane in the $y$-direction $s_{\mathrm{gap}}$, ending gap width to ground plane $e_{\mathrm{gap}}$ and end taper width $w_{\mathrm{e}}$=50nm. The nanopillar array is placed directly adjacent to the wide input of the taper. To avoid the demands of needing to model all elements of the nanoscale array plus a mm-long taper, we model the array as a single effective impedance $Z_{\mathrm{arr}}$ seen by the taper. The simulated input current density $J_{\mathrm{start}}$ is taken where it is strongest at the input (close to the taper edge), with the output density $J_{\mathrm{end}}$  sampled at the end of the taper. The total current $I_{\mathrm{start}}$  and $I_{\mathrm{end}}$ can also be calculated by integrating magnetic field $H$ around the taper input cross section. 

Figure \ref{fig:taper1} shows a plot of normalised peak current density coming from the array at the start of the taper $J_{\mathrm{start}}$ as a function of $s_{\mathrm{gap}}$ and taper length $t$. The peak in input current occurs when the taper reactance matches that of the array, with longer tapers requiring less of a change in CPW gap from start to end. The relation between taper reactance and taper length can be seen in Figure \ref{fig:taper2}, with an example plot of  Im($Z_{\mathrm{taper}}$) from the modeled geometry and kinetic inductance of the material with $s_{\mathrm{gap}}$=12$\mu$m. Starting highly capacitive, the taper smoothly becomes increasingly inductive as length is increased, with crossover reached around 2.1mm, which corresponds to the peak in input current $J_{\mathrm{start}}$ seen in Figure \ref{fig:taper1}. The peak current at the taper end $J_{\mathrm{end}}$ ($I_{\mathrm{end}}$) also corresponds to this matching. 

We note that for longer tapers ($t$$>$2.5mm) the reactance behaviour becomes much less smooth, with a discontinuity at 3.9mm. We attribute this to distributed reflection resonance inside the taper, forming a current antinode at the taper end. This is supported by Figure \ref{fig:taper3} shows a plot of the ratio of start to end to current density where the ratio spikes from around  $J_{\mathrm{start}} / J_{\mathrm{end}}$=0.1-0.5 at $t$$<$2.5mm to a factor of 3-4 at $t$=4mm. Additionally in Figure \ref{fig:taper4} we plot $J_{\mathrm{start}} / J_{\mathrm{end}}$ as a function of the impedance mismatch in the two regimes $t$$<$2.5mm and $t$$>$2.5mm, showing the different in behaviour between the smooth transformation and reflectance regimes. 

For $t$$<$2.5mm we achieve a stable, smooth impedance transformation but with a loss in the taper current $J_{\mathrm{start}} / J_{\mathrm{end}}$=0.1 even when close to impedance match. We consider that this loss in the taper could be reduced through better design e.g. using an exponential or Klopfenstein taper \cite{Klopfenstein1956}. Current transport in the taper can likely be improved by operating with longer taper in the reflection resonance regime, but this is likely to be more difficult to engineer. Through improvements, we consider it feasible that unity $J_{\mathrm{start}} / J_{\mathrm{end}}$ can be reached while retaining impedance match and high input current from the array.  

\section{NV Center Readout}

Using the current capture structure, IF current as a result of the axion electric field is focused into a single detection point, rather than  being distributed across the full area of the array and collector. For readout of this highly localised current at the taper end, an ensemble thin layer of NV centers in a small (111) diamond is employed \cite{Maze2008,Taylor2008,Doherty2013}, rotated such that the NV axis can read the primary vector field component from the current concentrated by the taper. NV centers are idea for this purpose, as their $m_s$=0 to $m_s$=1 spin transition can be aligned directly at 30.89GHz at 1T of applied magnetic field, can record from a microscopic volume in high proximity to the current (unlike a SQUID) and suffer fewer problems with mutual coupling to the waveguide structure. It is assumed for simplicity that all NVs are uniformly pointing in the [111] $z$-direction, offset to sample the maximum transverse circulating field from the current at the end of the taper. In a simple dipolar model, the approximate total magnetic field generated at each NV from a unit length of current carrying taper in the shorted end of the CPW can be estimated by:

\begin{equation}
B_{\mathrm{NV}} = \sum_{N} \frac{\mu_0 \, d I_{\mathrm{arr}}}{2\pi d_{\mathrm{nv}}}
\end{equation}

where $N$ is the emsemble size and $d_{\mathrm{nv}}$ is the depth of an NV center in the ensemble. The total field can be obtained by summing over all NVs and current slices. Total field $B_{\mathrm{NV}}$ can also be determined directly from simulation in Ansys for a given NV depth. 

The diamond is addressed with a 532nm laser, directed optically in the $z$-direction into the diamond. It is assumed it is possible to direct the laser and collect fluorescence along a similar optical path. NV readout can be performed using an electron paramagnetic resonance (EPR)-style approach, sending first a green laser pulse to initialise the ensemble into the ground $m_s$=0 state, followed by a $\pi$-pulse at $f_{\mathrm{IF}}$=30.89GHz using an antenna close to the diamond to initialise the ensemble in the high contrast $m_s$=+/-1. The system is then allowed to evolve freely with the transverse magnetic field from the axion induced signal at a frequency on resonance with the NV transition inducing (fractional) Rabi oscillation of the ensemble. This method ensures the NV control pulse does not act on the CPW while the NVs are capturing the axion signal. The degree of NV ensemble spin rotation in relaxation time T$_2^{*}$ is given by:

\begin{equation}
\theta = \gamma_e B_{\text{sig}} T_{2}^{*}
\end{equation}

and change in signal:

\begin{equation}
\Delta F = C N_{\mathrm{NV}} \theta N_{\mathrm{ph}}
\end{equation}

where $C$ is the maximum ensemble contrast ($\approx$ 0.005 for a small ensemble) $N_{\mathrm{NV}}$ the ensemble size and $N_{\mathrm{ph}}$ the detected fluorescence photon rate. 

\section{Noise Calculation}

%assumptions
%1. Dielectric constants remain fixed 
%2. Ignore conversion gain noise - variations in geometry
%3. Q remains fixed - good consistent CPW coupling
%4. Laser RIN is close to (if not at) the shot noise. 
%5. Laser polarisation is fixed

Potential noise sources are modelled in order to determine the overall readout time required to resolve the axion signal. As the current collector and NV act to simply collect and readout signal, the maximum possible axion signal results from the array, the collector and the NV acting only to reduce signal or add additional noise. 

\subsection{Johnson-Nyquist Noise}

The array nanogaps contribute both resistive and reactive contributions to impedance. The power spectral density of Johnson-Nyquist current noise $S_{I,\mathrm{JN}}$ in A$^{2}$/Hz , given by:

\begin{equation}
S_{I,JN}
 = 4 k_b T \mathrm{Re}(Y_{gap})
\end{equation}

where Re($Y_{\mathrm{gap}}$) is the real part (conductance) of the admittance of a single nanogap, from either numerical simulation or calculated analytically via the dielectric loss tangent. For the full nanogap array, we assume both current and thermal noise sum with the number of elements, $S_{I,\mathrm{JN}}^{\mathrm{arr}} = (M \times N)\, S_{I,\mathrm{JN}}^{\mathrm{gap}}$ as each gap contributes incoherent thermal noise to the overall array current. 

\subsection{LO Shot Noise}

The shot noise $S_{P,\mathrm{shot}}$ of the LO laser is given by:

\begin{equation}
S_{P,shot} = 2 h \omega_{LO} P_{av}
\end{equation}

where $P_{av}$ is the average laser power over a given bandwidth-defined integration time. This can be converted into equivalent electric field noise on $E_{LO}$ in (V/m)$^2$/Hz as:

%now correc,t matches code 27/6
\begin{equation}
S_{E,\text{shot}} = \frac{h \omega_{\text{LO}}}{2 n_{\text{opt}} c \varepsilon_0 A}
\end{equation}

where $n_{opt}$ is the optical refractive index of the mixing layer and A is the cross section area of the mixing region.
This gives a polarisation noise density of: 

%remove /4 due to definition of E field as peak
\begin{equation} S_{PL,\text{shot}} = \left( \varepsilon_0 \, n_{\text{opt}}^{4} \, r_{33} \, \left\vert{}E_{\text{LO}}\right\vert{} \right)^{2} S_{E,\text{shot}} \end{equation}

and a shot noise current density per nanogap of:

\begin{equation}
S_{J,shot}
=
\omega_{\mathrm{IF}}^{2}
\, S_{PL,shot}
\end{equation}

From the current noise density we calculate the total current noise as $S_{I,\mathrm{shot}} = A_{\mathrm{eff}}^{2}\, S_{J,\mathrm{shot}}$, where $A_{\mathrm{eff}}$ is the facing area of the pillars that forms the capacitance of the nanogap. This can be calculated analytically or derived from finite element simulation. As it is assumed each nanogap is an independent RF source, S$_{I,shot}^{\mathrm{arr}}$=MxN S$_{I,shot}^{\mathrm{gap}}$. 

\subsection{LO Phase Noise}

We calculate LO phase noise as:

\begin{equation}
S_{\phi}(f) = \frac{\Delta \nu}{\pi f_{\mathrm{offset}}^{2}}
\end{equation}

where $S_{\phi}$($f$) is the single-sided phase noise power spectral density, $\Delta \nu$ the FWHM laser linewidth and  $f_{\mathrm{offset}}$ the Fourier offset frequency from the carrier (Hz). We calculate current density noise as:

\begin{equation}
S_{J, phase} = |J|^{2}\, S_{\phi} 
\end{equation}

where $S_{I,\mathrm{phase}} = A_{\mathrm{eff}}^{2}\, S_{J,\mathrm{phase}}$. For phase noise, we take the total array noise to scale as $(M \times N)^2$.

\subsection{Dielectric Noise}

Dielectric noise can be calculated separately from our Johnson-Nyquist noise (for Ansys calculations this is included in admittance $Y_{\mathrm{gap}}$; care is taken not to double count noise sources) from the imaginary part of the dielectric susceptibility (loss factor) $\chi''$ , representing dielectric dissipation

\begin{equation}
S_{J,diel} =
\frac{4 k_B T \, \varepsilon_0 \, \chi'' \, \omega}{V}
\end{equation}

where $S_{I,\mathrm{diel}} = A_{\mathrm{eff}}^{2}\, S_{J,\mathrm{diel}}$ and scaling as $M \times N$ for the total array noise.

\subsection{LO Relative Intensity Noise}

For a balanced few-mW level at at 1550nm relative intensity noise (RIN) of -150 to -170dB/Hz can be achieved, close to the shot noise level, given by: 

\begin{equation}
\mathrm{RIN} = \frac{S_P}{P_0^2}
\end{equation}

where $S_P$ is the noise power spectral density at average laser power $P_{\mathrm{LO}}$ given by:

\begin{equation}
P_0 = \frac{1}{2} n \epsilon_0 c A |E|^2
\end{equation}

with $A$ is the laser spot area, relating intensity, power and electric field. This yields: 

\begin{equation}
S_{|E|^2} = \frac{4}{(n \epsilon_0 c A)^2} S_P
\end{equation}

Finally:

\begin{equation}
S_{P_{\mathrm{RF}}} = \left(\epsilon_0 n^4 r_{33} E_{\mathrm{ax}}\right)^2 \frac{1}{4 |E_{\mathrm{LO}}|^2} S_{|E_{\mathrm{opt}}|^2}
\end{equation}

\begin{equation}
S_{J,RIN} = \omega^2 S_{P_{\mathrm{RF}}}
\end{equation}

and $S_{I,\mathrm{RIN}} = A_{\mathrm{eff}}^{2}\, S_{J,\mathrm{RIN}}$. For RIN, the total array noise scales as $(M \times N)^2$. 

We note overall that laser frequency and phase stability is important, as any drift will be transmitted to the GHz IF, with signal lost due to mismatch. This is achievable using stabilised fibre or external cavity laser systems. It is assumed that slow drift in frequency and power can be reduced as much as possible. We do not evaluate effects hard to quantify such as backreflection or unwanted cavity modes. 

\subsection{NV Readout Noise}

For each NV center, noise arises from the center optical readout and spin dynamics and from conversion of the current noise $I_{\mathrm{noise}}$ of the axion signal source generated by the above processes. For a single NV considering a simple dipole model, the magnetic spectral noise density can be written as: 

\begin{equation}
S_B = \kappa^2 \, I_{\mathrm{noise}}^2
\end{equation}

where: 

\begin{equation}
\kappa = \frac{\mu_0}{2\pi r}
\end{equation}

with $S_B$ in T$^2$/Hz. A white noise spectrum is assumed over the NV integration bandwidth, such that the accumulated noise in time $T_{2}^{*}$ can be written as: 

\begin{equation}
\sigma_{\phi}^2 = \gamma_e^2 S_B T_2^*
\end{equation}

This gives a variation in the signal due to the predominantly array-generated current noise of:

\begin{equation}
\sigma_{\Delta F_{\mathrm{arr}}}^2 = \left(N_{\mathrm{ph}} C_{\max} N_{\mathrm{NV}}\right)^2 \sigma_{\phi}^2
\end{equation}

By the definition of $S_B$ and $\Delta F$, the Poission-limited shot noise variance for the NV readout is given by:

\begin{equation}
\sigma_{\Delta F_{\mathrm{NV}}}^2 = R_{\mathrm{ph}} T_2^* = N_{\mathrm{ph}}
\end{equation}

It is assumed that the NV readout can be configured to minimise technical noise, reaching close to the shot noise limit for a suitable ensemble size and coherence time. 

\section{196THz Sensitivity Estimates}

%Add here a table list of the basic parameters.
\begin{table}[h]
\centering
\begin{tabular}{lll}
\hline
Parameter & Value & Description \\
\hline
$m_a$ & $0.8$ & Axion mass (eV) \\
$g_{a\gamma\gamma}$ & $10^{-11}$ & Axion-photon coupling parameter \\
$B_0$ & $1$ & Magnetic field (T) \\
$T$ & $4.2$ & Temperature (K) \\
$n_{\mathrm{opt}}$ & $2.3$ & LiNb optical refractive index \\
$r_{eff}$ & $35$ & LiNb electro-optic coefficient (pm/V) \\
$\varepsilon_{\mathrm{Li}}$ & $28$ & LiNb 30GHz relative permittivity \\
$\tan \delta$ & $10^{-4}$ & Dielectric loss tangent \\
$\eta_{c}$ & 1 & Fill factor \\
$I_{start}$/I$_{end}$ & 1 & Taper current transfer ratio \\
$r$ & $60$ & Pillar radius (nm) \\
$h_{\mathrm{pillar}}$ & $120$ & Pillar height (nm) \\
$g$ & $5$ & Gap size (nm) \\
$N$ & $1000$ & Array elements (x-direction) \\
$M$ & $1000$ & Array elements (y-direction) \\
$N_{\mathrm{NV}}$ & $10^{4}$ & NV ensemble size centers \\
$C$ & $0.005$ & NV readout contrast \\
$T_2^*$ & $10^{-6}$ & NV coherence time (s) \\
\hline
\end{tabular}
\caption{Simulation parameters used in the model. Note that we model the upper limit achievable sensitivity, assuming all array current is collected by the taper $\eta_{c}$=1, $I_{start}$=$I_{arr}$.}
\label{tab:parameters}
\end{table}

\begin{figure}[htbp]
    \centering
    \includegraphics[width=0.5\textwidth]{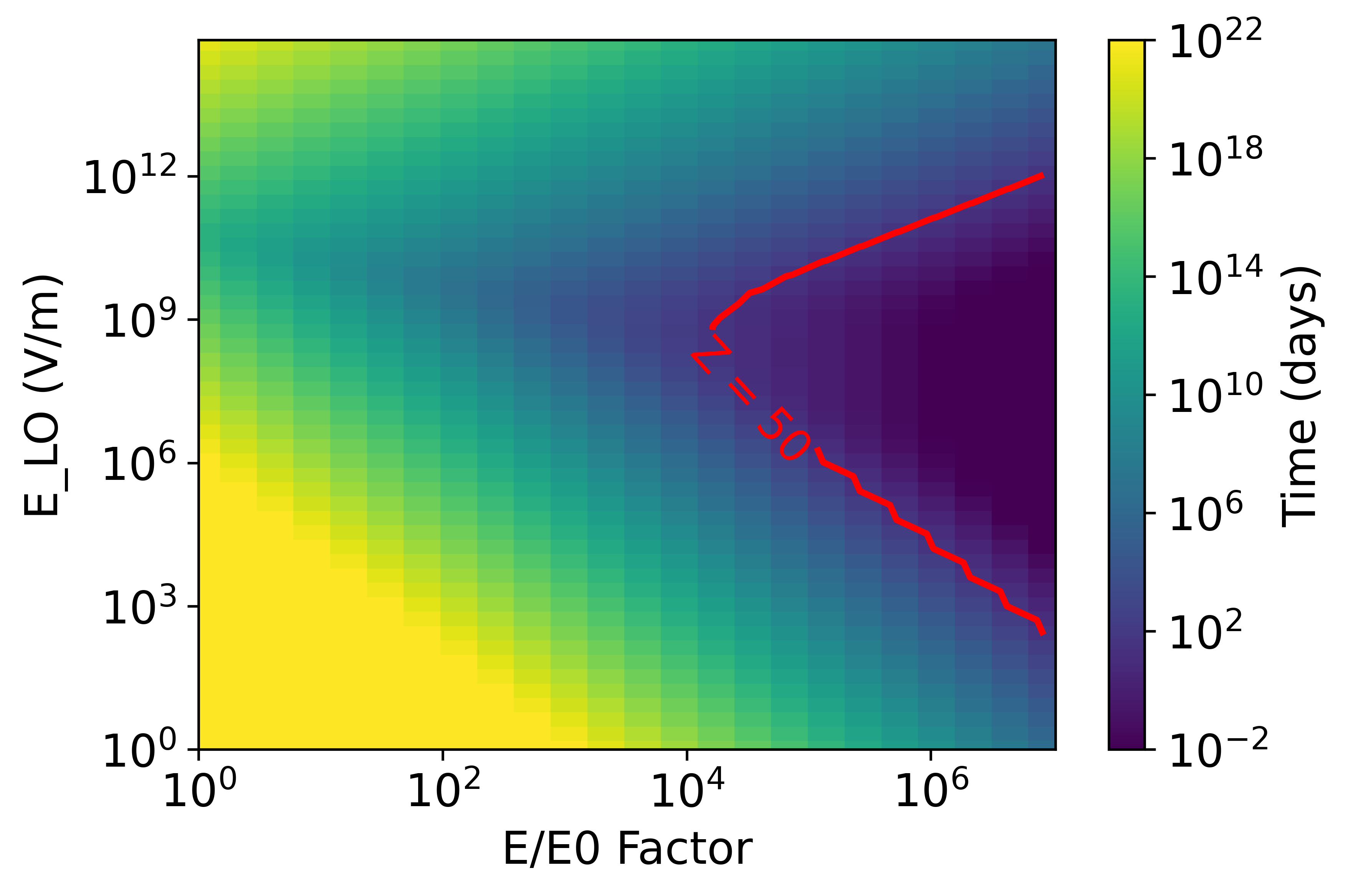}
    \caption{Time required for SNR=1 in days using only current data from the array as a function of applied laser local oscillator electric field $E_{\mathrm{LO}}$ and enhancement factor $E$/$E_{0}$. Indicated is the border of the $t_m$=50 day period for a reasonable experimental run.}
    \label{fig:figure2}
\end{figure}

\begin{figure}[htbp]
    \centering
    \includegraphics[width=0.48\textwidth]{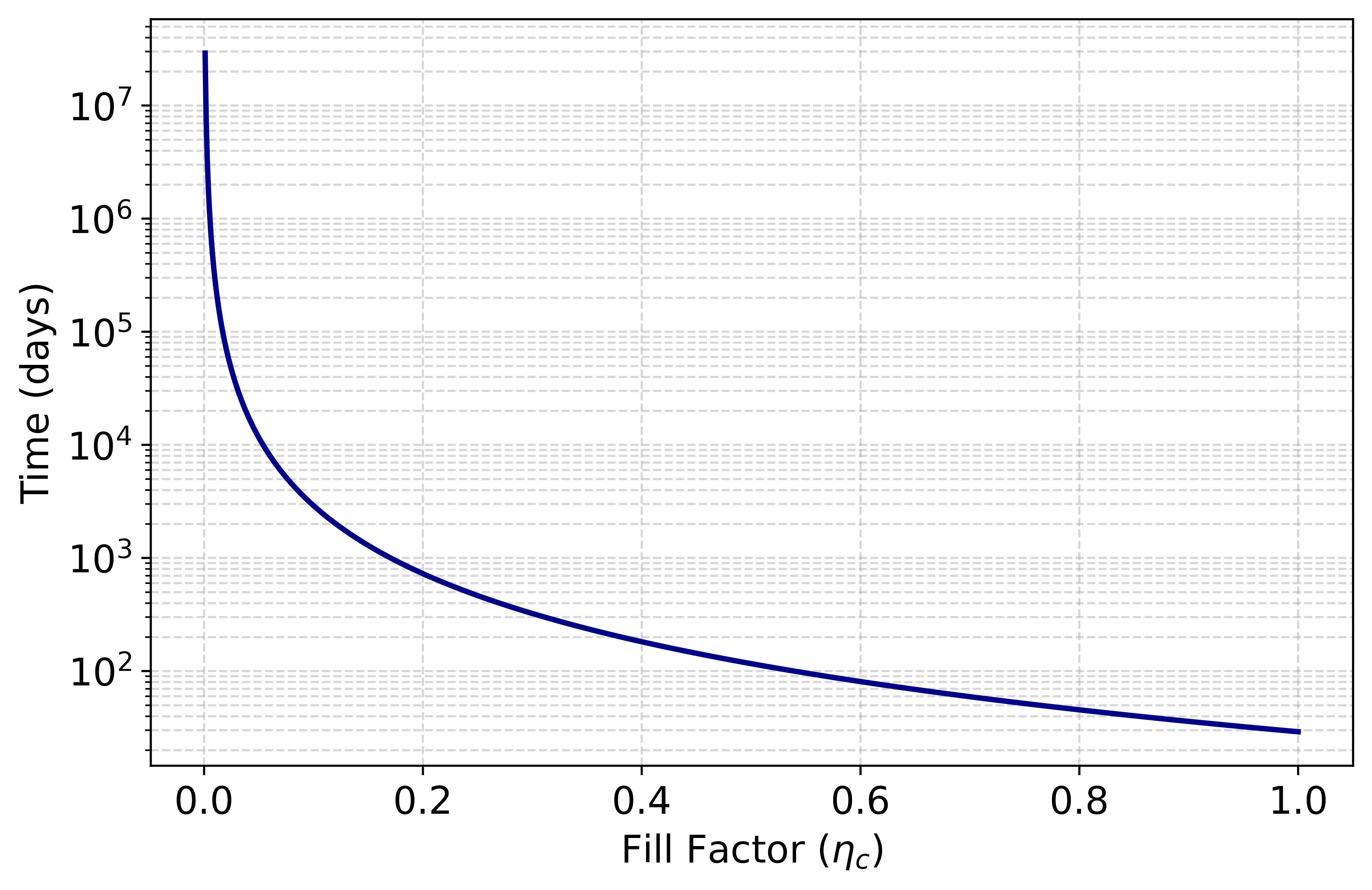}
    \caption{Time required for SNR=1 in days varying factor $\eta_c$ with $E_{\mathrm{LO}}$=10$^{6}$ and $E$/$E_{0}$=10$^{5}$ on the edge of the $t_m$=50 day window. Time required increases considerably for low $\eta_c$, if this is not compensated through waveguide engineering }
    \label{fig:etaplot}
\end{figure}

For 196THz/1550nm, calculations are performed at axion mass $m_{a}$$\approx$0.8eV giving a baseline $E_{\mathrm{ax}}$=1.625 $\times$10$^{-12}$ V/m. The spatial (axion de Broglie) coherence length is 1.55mm. The parameters used in the simulation model are given in Table \ref{tab:parameters}.

Unless otherwise stated, results presented in this section  make the assumption of ideal current transmission from array to taper, such that all array current reaches the end of the taper with $\eta_{c}$=1, $I_{\mathrm{start}}$=$I_{\mathrm{arr}}$. We assume the waveguide and taper can be engineered such as to overcome the reduction in signal associated with mode overlap. As such this work presents a best possible limit for signal resolution. The variation of time with $\eta_{c}$ can be seen in Figure \ref{fig:etaplot} with time dramatically increased for low values. 

The situation with zero electric field enhancement and a very small laser field $E_{\mathrm{LO}} = 1\,\mathrm{V/m}$ is first modelled. In this regime, the array noise is entirely dominated by J-N noise from the device structure itself, with $S_{I}^{\mathrm{JN}} = 6.625 \times 10^{20}\,\mathrm{A^2/Hz}$, $R = \mathrm{Re}(Z_{\mathrm{gap}}) = 3.49 \times 10^{9}\,\Omega$, $C_{\mathrm{gap}} = 16.2\,\mathrm{aF}$ and $M = N = 1000$. To achieve a SNR = 1 at this noise floor in a 50 day measurement run, a baseline signal level $I_{\mathrm{arr}} = 8.756 \times 10^{-14}\,\mathrm{A}$ is required for axion-induced current from the pillar array.

Figure \ref{fig:figure2} shows the signal integration time required for SNR=1 as a function of incident laser electric field E$_{LO}$ and an electric field enhancement factor acting on both $E_{\mathrm{ax}}$ and $E_{\mathrm{LO}}$, resulting from nanogap confinement. The calculation assumes CW laser illumination. A marked region at $E$/$E_0$$>$10$^{4}$ shows achievable measurement times of 50 days or less. Reaching this region depends on two factors: a) the applied laser power must be low enough to not damage the array or cause excessive heating and b) the enhancement factor must be realisable. For a standard cooling power of a cryostat at 4.2K an optimistic upper limit on power input at $P_{LO}$=1W can be taken, corresponding to $E_{\mathrm{LO}}$$\approx$5$\times$10$^{4}$V/m. To reach 50 days, this demands an extremely high enhancement factor of approximately $E$/$E_0$=10$^{6}$, which is likely not achievable by engineering the nanogap alone. 

\begin{figure}[htbp]
    \centering
    \includegraphics[width=0.5\textwidth]{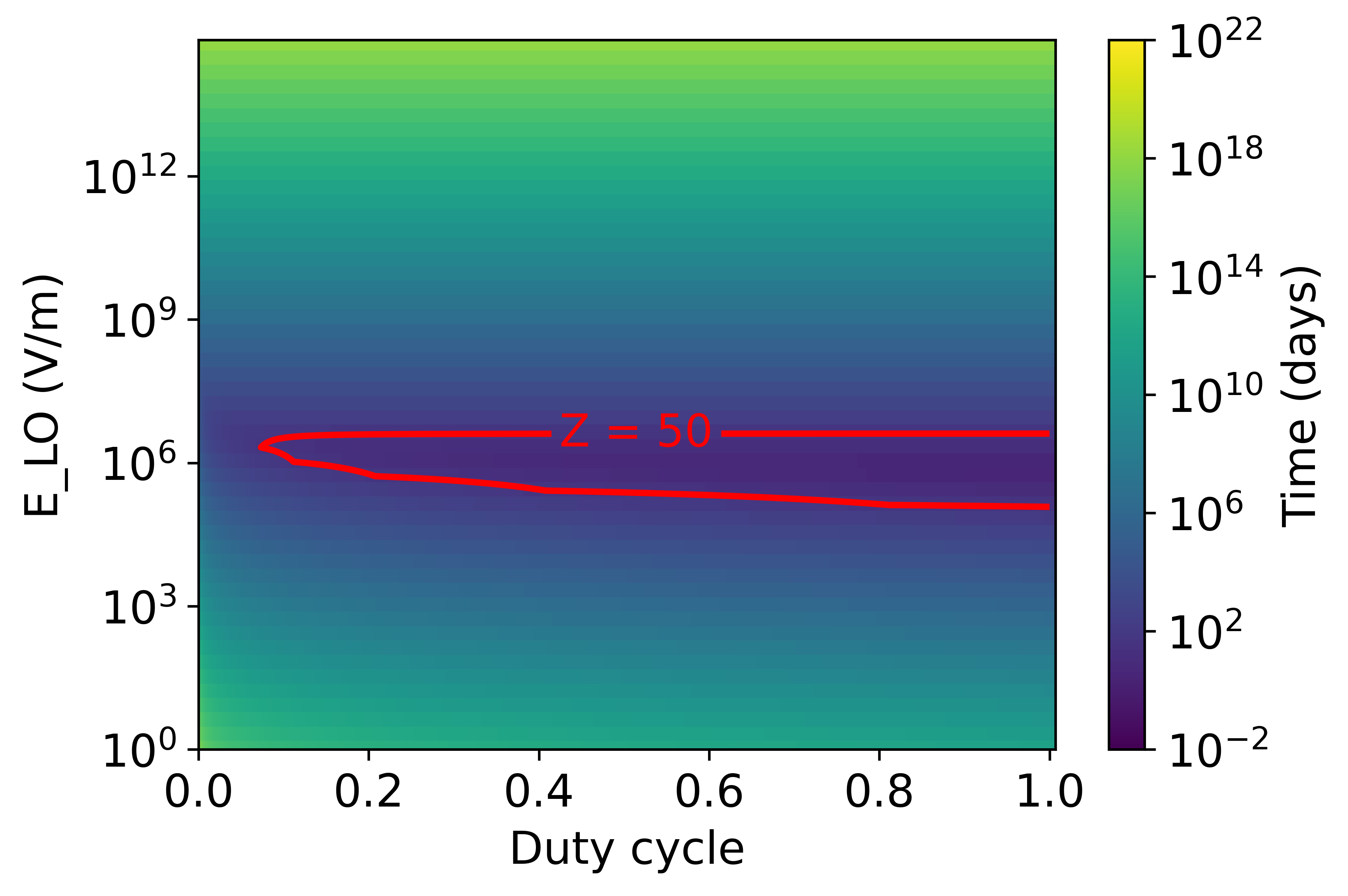}
    \caption{Time required for SNR=1 in days using only current data from the array as a function of applied laser local oscillator electric field $E_{\mathrm{LO}}$ laser duty cycle, for a fixed $E$/$E_0$ enhancement factor of 10$^{5}$ Indicated is the border of the $t_m$=50 day period for a reasonable experimental run.}
    \label{fig:figure3}
\end{figure}

The power issue can be addressed in two ways: by using a pulsed laser with a reduced duty cycle or by implementing an optical cavity above the nanopillar surface. The latter option would reduce the required input power to reach a given electric field by cavity factor $Q_{opt}$. By forming a cavity of $Q_{opt}$= 10$^{2}$-10$^{5}$ significant reduction in LO laser input power could be achieved, bringing more of the viable experimental region into play with a lower enhancement factor. For use of a pulsed laser we simulate measurement time as a function of duty cycle. Figure \ref{fig:figure3} shows a plot of duty cycle versus $E_{\mathrm{LO}}$. Fot the lowest duty cycle at the edge of the 50 day window, $E_{\mathrm{LO}}$$\approx$10$^{8}$ for a duty cycle of 0.03. For a 200$\mu$m spot size covering our array, this requires a peak power of 3.83MW and average power of 115kW, a value impractically high for realistic cooling. Although pulsing achieves higher peak $E_{\mathrm{LO}}$, the low duty cycle increases measurement dead time, limiting sensitivity.  

\begin{figure}[htbp]
    \centering
    \includegraphics[width=0.5\textwidth]{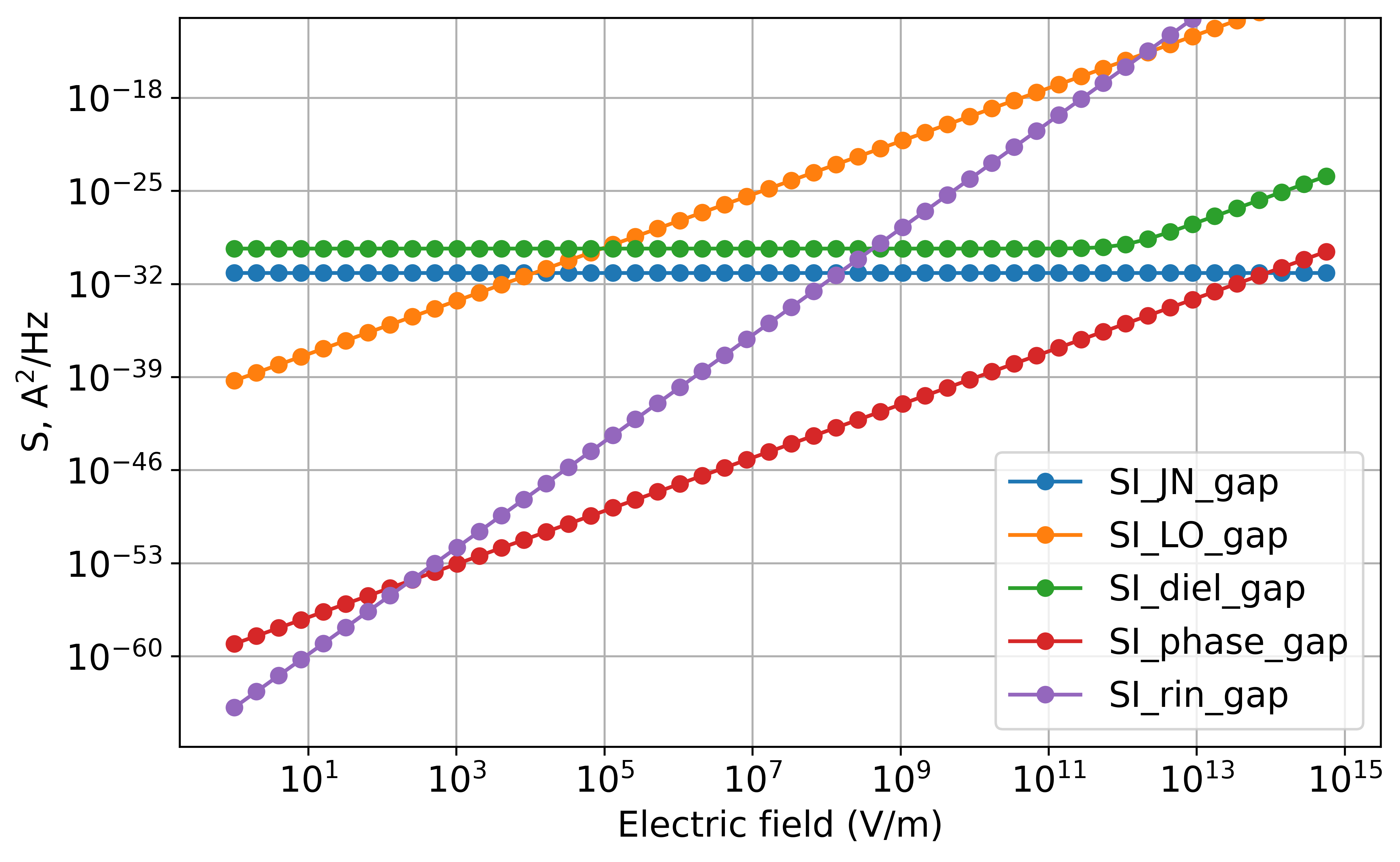}
    \caption{Power spectral density of all noise sources in A$^2$/Hz as a function of $E_{\mathrm{LO}}$, assuming a fixed enhancement factor of $E$/$E_0$=10$^{4}$. At low local oscillator power, noise is dominated by the nanogap thermal noise, the level of which is defined by unchanging structural parameters. At higher LO power, the LO shot noise begins to dominate.}
    \label{fig:nvn}
\end{figure}

In Figure \ref{fig:nvn} the noise spectral density for each type of noise is plotted. It can be noted that in the domain where 50 day or less experimental runs could resolve the axion that the primary limitations are thermal noise from the array at lower field values and local oscillator (shot) noise at higher LO power. 

\begin{figure}[htbp]
    \centering
    \includegraphics[width=0.48\textwidth]{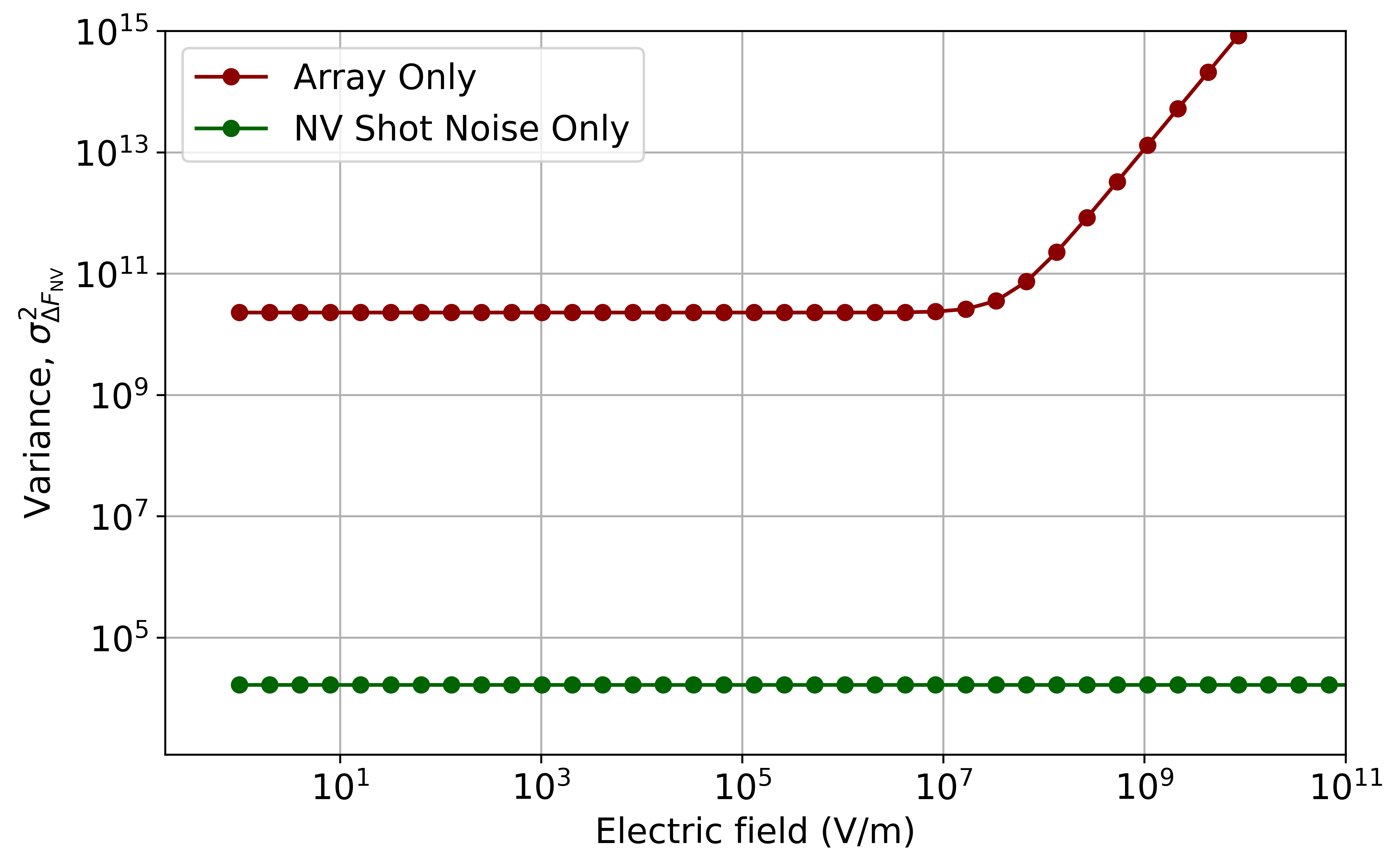}
    \caption{Variance in the NV readout signal due to the current noise coming from the nanopillar array and due to the intrinsic shot noise of the readout. Here typical NV ensemble parameters are modelled: $N_{NV}$=10$^{4}$, contrast $C$=5$\%$, $T_{2}^{*}$=1$\mu$s and operating at saturation with $P_{\mathrm{sat}}$=25mW and fluorescence power $P_{\mathrm{fl}}$=0.23$\mu$W. In this work, NV readout is overwhelmingly dominated by the noise in the axion signal current rather than the NV readout itself. }
    \label{fig:nvn}
\end{figure}

In Figure \ref{fig:nvn} a comparison is shown between the noise floor resulting from current noise predominantly from the array and intrinsic shot noise of the NV ensemble under achievable operating conditions for sensing. It can be seen that even for a simple approximation of the magnetic field, the noise of the NV sensor is far below the noise floor of the rest of the sensor. This is also true in the limit of realistic NV-related technical noise several orders of magnitude above the shot noise floor. The NV stage acts as a high quality, microscale probe of the signal output, without constraing axion sensing.

\section{24Thz Sensitivity Estimates}

We finally consider sensing at 24THz (equivalent axion mass $m_a$=0.1eV). At 24THz we can no longer use LiNb as the ectro-optic medium for DFG as the frequecy lies within the Reststrahlen bands of the material \cite{Cochard2017}, producing high optical absorption. We can instead switch to an alternative materials. Here we exemplify this for cadmium telluride (CdTe), with the following changes:

\begin{itemize}
    \item Per equation 5 we have the advantage of an immediate gain 1 order of magnitude in $E_{\mathrm{ax}}$=10$^{-10}$V/m to 10$^{-12}$V/m for coupling $g_{a\gamma\gamma}$=10$^{-10}$-10$^{-12}$. 
		\item Shorter wavelength requires wider and higher pillars and wider nanogaps, reducing M=N=200 for W$_{CPW}$=200$\mu$m. However, the de Broglie wavelength is increased to 1.3cm at 24THz, allowing a larger scale array to be used. 
    \item CdTe has a weaker electro-optic coefficient for DFG r$_{eff}$5$\times$10$^{-12}$m/V, reducing the IF signal.
		\item CdTe has sligthly larger n$_{opt}$=2.67, increasing the IF signal.
		\item CdTe has a smaller refractive index at IF $\epsilon_{r}$=10.2, reducing reactance and increasing signal. 
		\item CdTe has a lower dielectric loss $\tan \delta$=10$^{-4}$, reducing $R_{\mathrm{gap}}$ and Johnson-Nyquist noise $S_{I,\mathrm{JN}}$. 
		
\end{itemize}
		
\begin{figure}[htbp]
    \centering
    \includegraphics[width=0.5\textwidth]{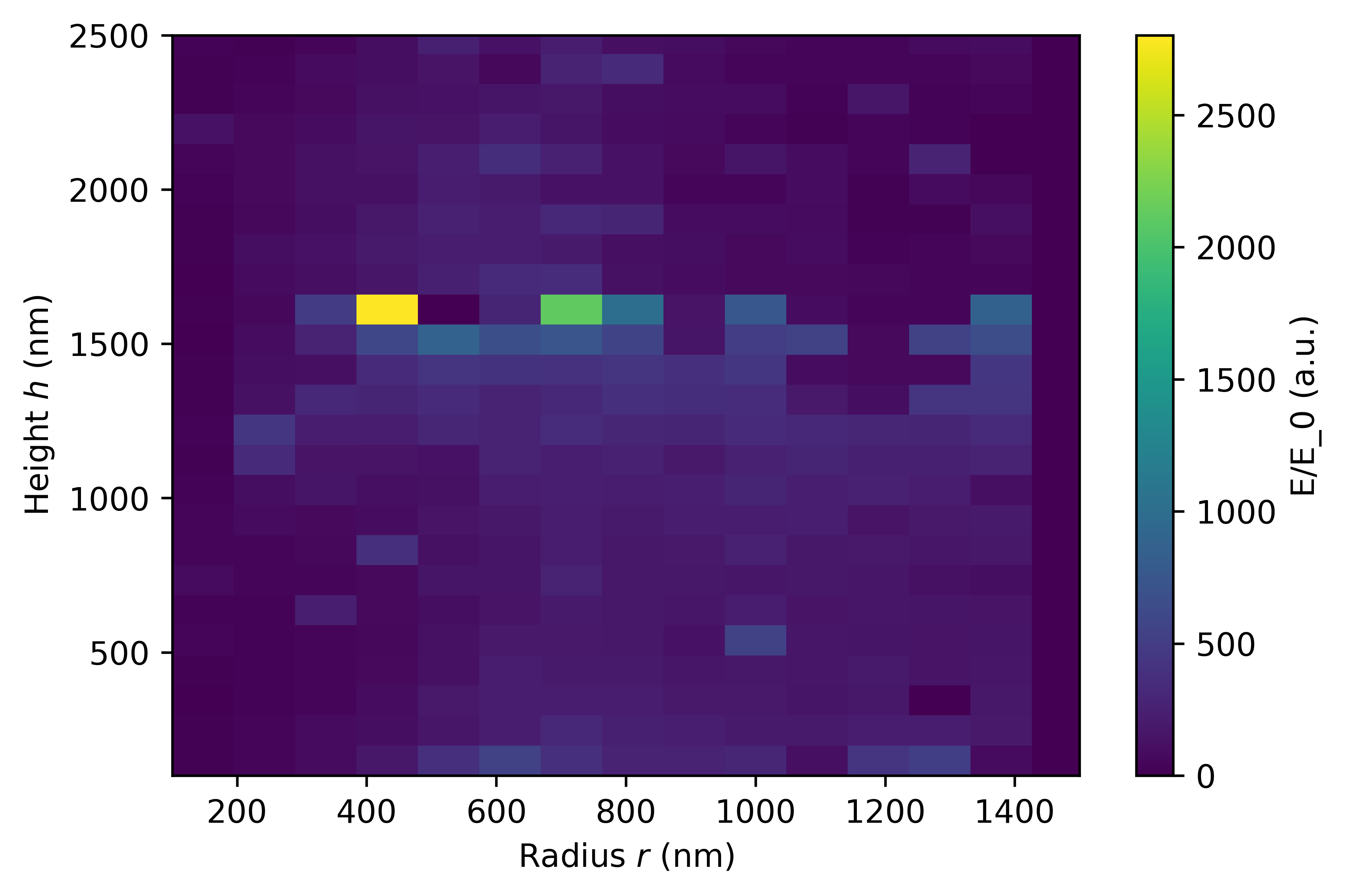}
    \caption{Plot of pillar radius versus height for CdTe, modelling a gap with of 5nm. A maximum enhancement factor of 2801.2 is simulated at $r$=400nm, height $h$=1600nm.}
    \label{fig:cdtepillars}
\end{figure}	

\begin{figure}[htbp]
    \centering
    \includegraphics[width=0.5\textwidth]{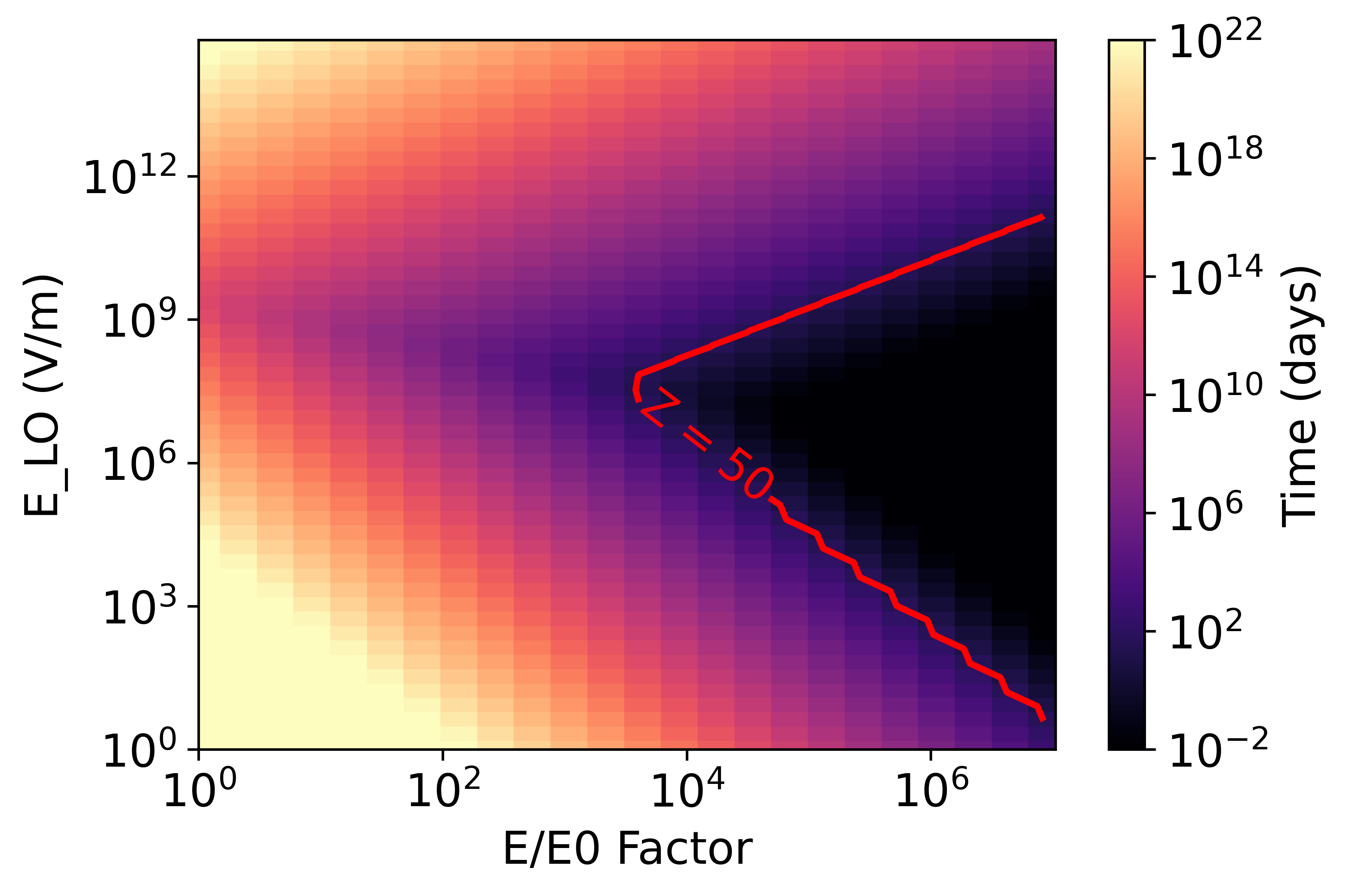}
    \caption{Time required for SNR=1 in days using CdTe in place of LiNb at 24THz as a function of applied laser local oscillator electric field $E_{\mathrm{LO}}$ and enhancement factor $E$/$E_{0}$. Indicated is the border of the $t_m$=50 day period for a reasonable experimental run. CdTe as simulated by the same methodology as for LiNb presents a larger viable window for investigation.}
    \label{fig:heatmapcdte}
\end{figure}	
		
Figure \ref{fig:cdtepillars} shows the simulation of enhancement factor as a function of pillar radius and height in Ansys, usig the same approach as for LiNb. Using the maximum $r$=400, $h$=1600nm values here, we simulate the time required for unity SNR for CdTe. The result can be seen in Figure \ref{fig:heatmapcdte} with CdTe offering improved performance over LiNb at 24THz and a larger viable experimental window.

\section{Discussion and Conclusions}

This work explores the possibility of developing a sensor consisting of a combination of THz frequency transducer with an NV quantum sensing to probe for axions of mid-range mass 10meV to 1eV. The viability of the scheme is first modelled for $m_a$$\approx$0.8eV ($\approx$196THz, 1550nm) using LiNb, a well established material for optics and electro-optic conversion schemes, at telecommunications (1550nm) wavelength. Based on assumptions made, a viable 50 day experimental time window is reached using an array of $M$=$N$=1000 nanogaps. We then simulate at 24THz for $m_a$$\approx$0.1eV, switching to CdTe, again showing a viable measurement time window. 

Several broad conclusions can be drawn from the investigation. First, it is challenging to achieve additional resonant signal enhancement once the signal is transduced into the microwave domain. Based on assumptions made, the simulated nanopillar array acts as a parallel high admittance sheet current source that when strongly coupled into a microwave resonator acts to reduce the Q factor significantly. It may be possible to design a weakly coupled scheme that avoids this, for example coupling into a mode of a cooled dielectric puck resonator \cite{Driscoll1992,McAllister2018}, but likely with the loss of too much of the ultra-weak axion-induced signal. These aspects mean that a superconducting waveguide current collector is likely required. Any normal state waveguide considered adds a high Johnson-Nyquist noise floor, rendering the axion signal undetectable. A superconducting current collector made from Type II material (e.g. Nb, NbTiN) must be carefully aligned in plane, parallel to the high applied magnetic field.

Further simulation work is required to fully simulate the elctrical properties of the pillar array and the current transfer from it, with fewer assumptions than necessary here. The treatment of the array as an admittance sheet and current source and as a single input to the tapered collector is an approximate simplification. In reality, each nanogap element will have a different coupling to the taper based on separation, position, phase difference and local EM fields. The structural fill factor $\eta_{c}$ used here is a simplification of the true modal overlap between the array gap emitters and the collector waveguide. These simplification is necessary to solve the problem with available computational resources; a full finite element simulation of field enhancement for thousands of nanoscale elements, optoelectronic mixing plus a mm-scale taper is beyond available capabilities and remains an important next step for this type of scheme. It is very possible that through careful design beyond the simple scheme here that a as a solution with $\eta_{c}$$\rightarrow$1 while maintaining high current collection and trasmission efficiency can be realised. 

The requirement for high heterodyne mixing electric field and the need for superconducting collection puts strict constraints on operation: a vertical (in z) Fabry–Pérot–type optical cavity using a top reflector layer is almost certainly required to keep local oscillator input power low enough to avoid heating. However, the design requirements on the optical cavity (especially Q-factor) are less strict than if the cavity were used for direct THz detection of e.g. the effect of the axion on refractive index, being required only to deliver optical power into the sensor. The cavity can be constructed out of plane avoiding strong dispersion effects in long path lengths in LiNb. 

The difficulty of realising resonant enhancement in the microwave domain means the majority of signal gain must come in the optical domain transduction stage. Simulations show that both heterodyne mixing and plasmonic enhancement are required to bring the axion signal up to a detectable level; neither is sufficient alone. The most simple method of enhancement using cylindrical nanopillars we simulate here produces enhancement factors in the 10$^{2}$ to 10$^{3}$ range, short of what is required for a viable axion signal measurement time. More advanced methods such as replacing the metal pillars with layered dielectric materials \cite{Kuznetsov2016} and surface lattice resonance structures \cite{Auguie2008,Maier2007} could help provide the extra degree of electric field enhancement required. The approach here ultimately trades absolute enhancement factor for interaction volume; the tighter the nanogaps can be fitted into the coherence volume of the axion, for example through 3D structuring, the stronger the signal.

\begin{figure}[htbp]
    \centering
    \includegraphics[width=0.5\textwidth]{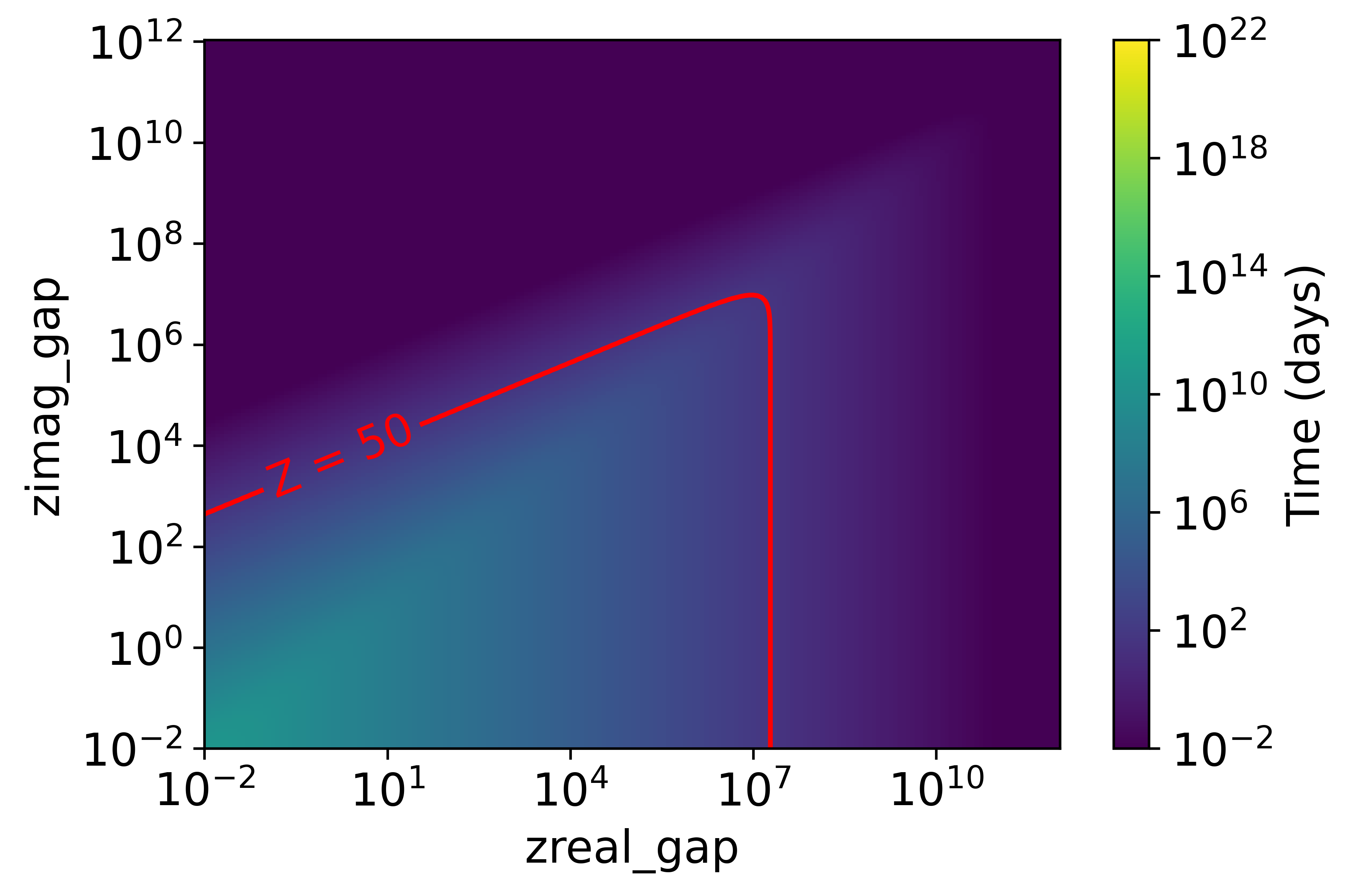}
    \caption{Time required as a function of the resistive and reactive parts of a single nanogap impedance for CdTe. The region requring more than 50 days is masked in red.}
    \label{fig:heatmap1cdtezarr}
\end{figure}	 

\begin{figure}[htbp]
    \centering
    \includegraphics[width=0.48\textwidth]{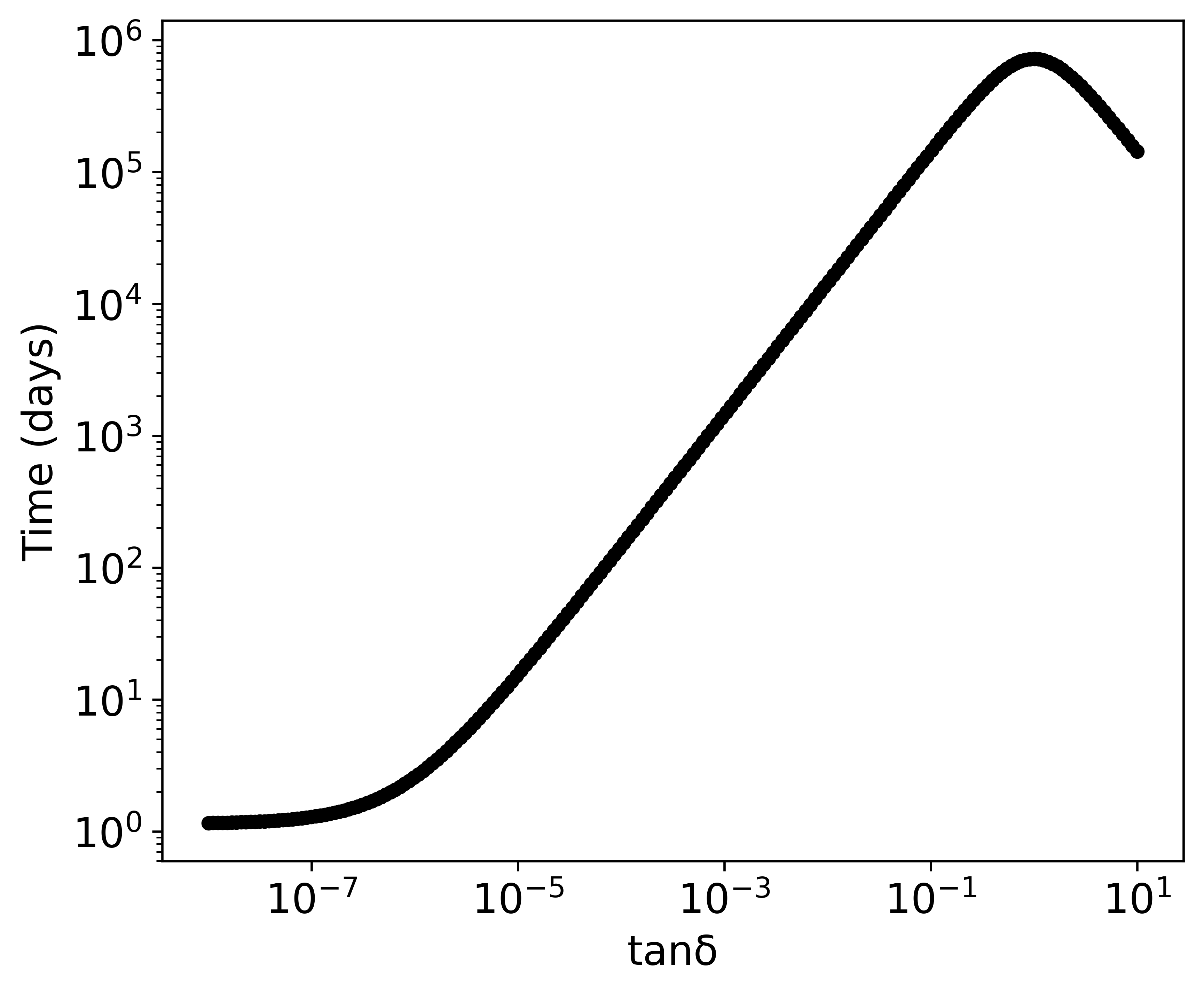}
    \caption{Time required for SNR=1 as a function of the dielectric loss tangent $\tan \delta$}
    \label{fig:tand}
\end{figure}	 

An important aspect not considered here is the effect of additional conductance in the nanogap e.g. from direct quantum tunneling, field emission (Fowler–Nordheim) tunnelling or photoinduced tunnelling that are harder to quantify. Instead in Figure \ref{fig:heatmap1cdtezarr} we show the viable experimental region for LiNb (Z$>$50 days) as a generalised function of the resistive and reactive components of a single gap and in Figure \ref{fig:tand} the time versus the loss tangent of the dielectric in the gap. Any additional contribution to admittance from e.g. tunneling can be included in these factors. We note that we need to operate in the regime of high resistance for high electric field enhancement factor ($>$10$^{7}$ Ohms for CdTe) at pillar size that gives a high reactance (>10$^{4}$). Sensor performance is ultimately bounded strongly by the dielectric quality and losses of the electro-optic layer. Pillar size and capacitance can be adjusted, but if $\tan \delta$ and conductance are too high then no viable experimental window is possible. It can be noted that tunneling rate could be used as the transduction mechanism itself, effectively acting in a similar manner to the optoelectronic method chosen for simulations \cite{Parzefall2015,Svetovoy2014}. Further simulation is also required to study competing factors (e.g. photothermal expansion, pyroelectric effects, photoelastic waves) that could have a component at the IF that drowns out the weak axion signal. 

Finally, a disadvantage of the scheme is relatively low tunability for a single array, the signal dropping by a factor of 5 if the axion field oscillation is 10 THz away from the peak enhancement set by the metasurface structure. However, it is very possible to rapidly nanofabricate multiple arrays with different structures (and hence operational frequencies) on a single multi-inch wafer. Both LO laser and the NV readout laser  can be readily relocated to these alternative arrays. Multiple small diamonds could be used for NV readout of each array, or developments in wafer scale diamond growth could allow the diamond to be used as the substrate \cite{Nelz2018WaferDiamond}. Reading out multiple of these arrays simultaneously splitting a single laser source could also be beneficial, for higher SNR, performing a gradiometer-style experiment for common mode noise cancellation and for investigating the spatial variance of the axion field. 

\section{Acknowledgements}

This work was funded by Villum Experiment Grant 57929 AXQM from the Villum Foundation.


\begin{thebibliography}{}

\end{thebibliography}


\begin{thebibliography}{10}

\bibitem{Wilczek1978}
Frank Wilczek.
\newblock Problem of strong $p$ and $t$ invariance in the presence of
  instantons.
\newblock {\em Physical Review Letters}, 40(5):279--282, 1978.

\bibitem{Peccei1977}
R.~D. Peccei and Helen~R. Quinn.
\newblock {$CP$ Conservation in the Presence of Pseudoparticles}.
\newblock {\em Physical Review Letters}, 38(25):1440--1443, June 1977.

\bibitem{Peccei1977b}
R.~D. Peccei and Helen~R. Quinn.
\newblock Constraints imposed by {$CP$} conservation in the presence of
  pseudoparticles.
\newblock {\em Physical Review D}, 16(6):1791--1797, September 1977.

\bibitem{Wilczek1987}
Frank Wilczek.
\newblock Two applications of axion electrodynamics.
\newblock {\em Physical Review Letters}, 58(18):1799–1802, May 1987.

\bibitem{Sikivie1983}
P.~Sikivie.
\newblock Experimental tests of the “invisible” axion.
\newblock {\em Physical Review Letters}, 51(16):1415–1417, October 1983.

\bibitem{Marsh2016}
David~J.E. Marsh.
\newblock Axion cosmology.
\newblock {\em Physics Reports}, 643:1–79, 2016.

\bibitem{Sikivie1985}
P.~Sikivie.
\newblock Detection rates for ‘‘invisible’’-axion searches.
\newblock {\em Physical Review D}, 32(11):2988–2991, December 1985.

\bibitem{Rybka2014}
Gray Rybka.
\newblock Direct detection searches for axion dark matter.
\newblock {\em Physics of the Dark Universe}, 4:14–16, 2014.

\bibitem{Bartram2021}
C.~Bartram et~al.
\newblock Search for invisible axion dark matter in the 3.3--4.2 $\mu$ev mass
  range.
\newblock {\em Physical Review Letters}, 127(26):261803, December 2021.

\bibitem{VanBibber1987}
K.~Van~Bibber, N.~R. Dagdeviren, S.~E. Koonin, A.~K. Kerman, and H.~N. Nelson.
\newblock Proposed experiment to produce and detect light pseudoscalars.
\newblock {\em Physical Review Letters}, 59(7):759–762, August 1987.

\bibitem{Schneider1984}
M.~B. Schneider, F.~P. Calaprice, A.~L. Hallin, D.~W. MacArthur, and D.~F.
  Schreiber.
\newblock Limit on im$(c_s c_a^*)$ from a test of $t$ invariance in $^{19}$ne
  beta decay.
\newblock {\em Physical Review Letters}, 52(8):695--695, February 1984.

\bibitem{Raffelt2024}
G.~Raffelt and A.~Caputo.
\newblock Astrophysical axion bounds: The 2024 edition.
\newblock In {\em Proceedings of 1st General Meeting and 1st Training School of
  the COST Action COSMIC WSIPers — PoS(COSMICWISPers)}, COSMICWISPers, page
  041. Sissa Medialab, March 2024.

\bibitem{aaaddd}
Ciaran O'HARE.
\newblock cajohare/axionlimits: Axionlimits, 2020.

\bibitem{Graham2011}
Peter~W. Graham and Surjeet Rajendran.
\newblock Axion dark matter detection with cold molecules.
\newblock {\em Physical Review D}, 84(5):055013, September 2011.

\bibitem{Kim1979}
Jihn~E. Kim.
\newblock Weak-interaction singlet and strong ${CP}$ invariance.
\newblock {\em Physical Review Letters}, 43:103--107, 1979.

\bibitem{Shifman1980}
M.A. Shifman, A.I. Vainshtein, and V.I. Zakharov.
\newblock Can confinement ensure natural ${CP}$ invariance of strong
  interactions?
\newblock {\em Nuclear Physics B}, 166(3):493--506, 1980.

\bibitem{Roy2023JWSTAxion}
Saptak Roy, Carlos Blanco, Christopher Dessert, Anirudh Prabhu, and Tea Temim.
\newblock Sensitivity of jwst to ev-scale decaying axion dark matter.
\newblock 2023.

\bibitem{Caldwell2017}
Allen Caldwell, Gia Dvali, Béla Majorovits, Alexander Millar, Georg Raffelt,
  Javier Redondo, Olaf Reimann, Frank Simon, and Frank~and Steffen.
\newblock Dielectric haloscopes: A new way to detect axion dark matter.
\newblock {\em Physical Review Letters}, 118(9), March 2017.

\bibitem{Baryakhtar2018}
Masha Baryakhtar, Robert Lasenby, and Jessie Teo.
\newblock Axion and hidden photon dark matter detection with multilayer optical
  haloscopes.
\newblock {\em Phys. Rev. D}, 98:035006, 2018.

\bibitem{Liu2021BREAD}
Jesse Liu, Kristin Dona, Gabe Hoshino, Stefan Knirck, Noah Kurinsky, Matthew
  Malaker, David Miller, Andrew Sonnenschein, Peter Barry, Karl~K. Berggren,
  et~al.
\newblock Broadband solenoidal haloscope for terahertz axion detection.
\newblock 2021.

\bibitem{Sekine2021}
Akihiko Sekine and Kentaro Nomura.
\newblock Axion electrodynamics in topological materials.
\newblock {\em Journal of Applied Physics}, 129(14), April 2021.

\bibitem{Lee2022TRAX}
Chang Lee and Olaf Reimann.
\newblock T-rax: Transversely resonant axion experiment.
\newblock {\em Journal of Cosmology and Astroparticle Physics}, 2022(09):007,
  2022.

\bibitem{K2020PhD}
Katrine Domino.
\newblock {\em New Advances in Transition Metal Catalyzed Reactions using
  Gaseous- and Fluorinated Building Blocks}.
\newblock PhD thesis, Aarhus University, Aarhus, Denmark, April 2020.
\newblock PhD dissertation, Interdisciplinary Nanoscience Center (iNANO).

\bibitem{green2025doubleresonancestrategyinterferometric}
Spencer Green and Frank Wilczek.
\newblock Double resonance strategy for interferometric detection of axions.
\newblock 2025.

\bibitem{McMahon2011}
Jeffrey~M. McMahon, Stephen~K. Gray, and George~C. Schatz.
\newblock Fundamental behavior of electric field enhancements in the gaps
  between closely spaced nanostructures.
\newblock {\em Physical Review B}, 83(11), March 2011.

\bibitem{Ward2010}
Daniel~R. Ward, Falco H\"{u}ser, Fabian Pauly, Juan~Carlos Cuevas, and Douglas
  Natelson.
\newblock Optical rectification and field enhancement in a plasmonic nanogap.
\newblock {\em Nature Nanotechnology}, 5(10):732–736, 2010.

\bibitem{GarcaMartn2011}
A.~García-Martín, D.~R. Ward, D.~Natelson, and J.~C. Cuevas.
\newblock Field enhancement in subnanometer metallic gaps.
\newblock {\em Physical Review B}, 83(19), May 2011.

\bibitem{Baumberg2019}
Jeremy~J. Baumberg, Javier Aizpurua, Maiken~H. Mikkelsen, and David~R. Smith.
\newblock Extreme nanophotonics from ultrathin metallic gaps.
\newblock {\em Nature Materials}, 18(7):668–678, April 2019.

\bibitem{Zong2019}
Cheng Zong, Ranjith Premasiri, Haonan Lin, Yimin Huang, Chi Zhang, Chen Yang,
  Bin Ren, Lawrence~D. Ziegler, and Ji-Xin Cheng.
\newblock Plasmon-enhanced stimulated raman scattering microscopy with
  single-molecule detection sensitivity.
\newblock {\em Nature Communications}, 10(1), November 2019.

\bibitem{Maier2007Plasmonics}
Stefan~A. Maier.
\newblock {\em Plasmonics: Fundamentals and Applications}.
\newblock Springer, 2007.

\bibitem{Martin2017}
Eamonn Martin, Colm Browning, Liam Barry, Arman Farhang, Linda Doyle, Manh~Ha
  Hoang, Matthias John, and Max Ammann.
\newblock 28 ghz 5g radio over fibre using uf-ofdm with optical heterodyning.
\newblock In {\em 2017 International Topical Meeting on Microwave Photonics
  (MWP)}, page 1–4. IEEE, October 2017.

\bibitem{10521695}
Farzaneh~A. Juneghani, Milad~G. Vazimali, Kim~F. Lee, Ectis Velazquez, Xun
  Gong, Gregory~S. Kanter, and Sasan Fathpour.
\newblock Wireless microwave-to-optical conversion on thin-film lithium
  niobate.
\newblock {\em Journal of Lightwave Technology}, 42(16):5583--5590, 2024.

\bibitem{Yoshioka:24}
Valerie Yoshioka, Jicheng Jin, and Bo~Zhen.
\newblock Coherent fir/thz wave generation and steering via surface-emitting
  thin film lithium niobate waveguides.
\newblock {\em Opt. Express}, 32(1):639--651, Jan 2024.

\bibitem{boyd2020nonlinear}
Robert~W. Boyd.
\newblock {\em Nonlinear Optics}.
\newblock Academic Press, Cambridge, MA, USA, 4th edition, 2020.

\bibitem{abel2019large}
Stefan Abel, Felix Eltes, Juerg~E. Baer, Folkert Horst, Daniele Caimi, Lukas
  Bedner, Marilyne Podpod, Radu Grigorescu, Chiara Marchiori, and Jean
  Fompeyrine.
\newblock Large pockels effect in micro-structured barium titanate thin films
  on silicon.
\newblock {\em Nature Materials}, 18(1):42--47, 2019.

\bibitem{zhou1995microwave}
X.~Zhou, T.~M. Crawford, S.~S. Eaton, and J.~Moreland.
\newblock Microwave dielectric properties of lithium niobate at cryogenic
  temperatures.
\newblock {\em Journal of Applied Physics}, 77(12):6471--6475, 1995.

\bibitem{tagantsev2003intrinsic}
Alexander~K. Tagantsev, Vladimir~O. Sherman, Konstantin~F. Astafiev,
  J.~Venkatesh, and Nava Setter.
\newblock Ferroelectric materials for microwave tunable applications.
\newblock {\em Journal of Electroceramics}, 11(1):5--66, 2003.

\bibitem{li2013probing}
Yilei Li, Yi~Rao, Kin~Fai Mak, Yumeng You, Shuyuan Wang, Cory~R. Dean, and
  Tony~F. Heinz.
\newblock Probing symmetry properties of few-layer mos2 and h-bn by optical
  second-harmonic generation.
\newblock {\em Nano Letters}, 13(7):3329--3333, 2013.

\bibitem{janisch2014extraordinary}
Corey Janisch, Haomin Wang, Wonho Cho, Talat~S. Rahman, and Zhiwen Liu.
\newblock Extraordinary second-harmonic generation in tungsten disulfide
  monolayers.
\newblock {\em Scientific Reports}, 4:5530, 2014.

\bibitem{Ebadi_2024}
Reza Ebadi, David~E. Kaplan, Surjeet Rajendran, and Ronald~L. Walsworth.
\newblock Galileo: Galactic axion laser interferometer leveraging
  electro-optics.
\newblock {\em Physical Review Letters}, 132(10), March 2024.

\bibitem{Davoudiasl2026}
Hooman Davoudiasl and Yannis~K. Semertzidis.
\newblock Cosmic axions revealed via amplified modulation of the ellipticity of
  a laser.
\newblock {\em Physical Review D}, 113(3), February 2026.

\bibitem{Marchiori2022}
E.~Marchiori, L.~Ceccarelli, N.~Rossi, G.~Romagnoli, J.~Herrmann, J.-C. Besse,
  S.~Krinner, A.~Wallraff, and M.~Poggio.
\newblock Magnetic imaging of superconducting qubit devices with scanning
  squid-on-tip.
\newblock {\em Applied Physics Letters}, 121(5), August 2022.

\bibitem{Beck2016}
M.~A. Beck, J.~A. Isaacs, D.~Booth, J.~D. Pritchard, M.~Saffman, and
  R.~McDermott.
\newblock Optimized coplanar waveguide resonators for a superconductor–atom
  interface.
\newblock {\em Applied Physics Letters}, 109(9), August 2016.

\bibitem{Klopfenstein1956}
R.~W. Klopfenstein.
\newblock A transmission line taper of improved design.
\newblock {\em Proceedings of the IRE}, 44(1):31--35, 1956.

\bibitem{Maze2008}
J.~R. Maze, P.~L. Stanwix, J.~S. Hodges, S.~Hong, J.~M. Taylor, P.~Cappellaro,
  L.~Jiang, M.~V.~Gurudev Dutt, E.~Togan, A.~S. Zibrov, A.~Yacoby, R.~L.
  Walsworth, and M.~D. Lukin.
\newblock Nanoscale magnetic sensing with an individual electronic spin in
  diamond.
\newblock {\em Nature}, 455(7213):644–647, October 2008.

\bibitem{Taylor2008}
J.~M. Taylor, P.~Cappellaro, L.~Childress, L.~Jiang, D.~Budker, P.~R. Hemmer,
  A.~Yacoby, R.~Walsworth, and M.~D. Lukin.
\newblock High-sensitivity diamond magnetometer with nanoscale resolution.
\newblock {\em Nature Physics}, 4(10):810–816, 2008.

\bibitem{Doherty2013}
Marcus~W. Doherty, Neil~B. Manson, Paul Delaney, Fedor Jelezko, J\"{o}rg
  Wrachtrup, and Lloyd~C.L. Hollenberg.
\newblock The nitrogen-vacancy colour centre in diamond.
\newblock {\em Physics Reports}, 528(1):1–45, 2013.

\bibitem{Cochard2017}
Charlotte Cochard, Thiemo Spielmann, Naoufal Bahlawane, Alexei Halpin, and
  Torsten Granzow.
\newblock Broadband characterization of congruent lithium niobate from mhz to
  optical frequencies.
\newblock {\em Journal of Physics D: Applied Physics}, 50(36):36LT01, August
  2017.

\bibitem{Driscoll1992}
M.M. Driscoll, J.T. Haynes, R.A. Jelen, R.W. Weinert, J.R. Gavaler,
  J.~Talvacchio, G.R. Wagner, K.A. Zaki, and X.-P. Liang.
\newblock Cooled, ultrahigh q, sapphire dielectric resonators for low-noise,
  microwave signal generation.
\newblock {\em IEEE Transactions on Ultrasonics, Ferroelectrics and Frequency
  Control}, 39(3):405–411, May 1992.

\bibitem{McAllister2018}
Ben~T. McAllister, Graeme Flower, Lucas~E. Tobar, and Michael~E. Tobar.
\newblock Tunable supermode dielectric resonators for axion dark-matter
  haloscopes.
\newblock {\em Physical Review Applied}, 9(1), January 2018.

\bibitem{Kuznetsov2016}
Alexander~I. Kuznetsov, Andrey~E. Miroshnichenko, Mark~L. Brongersma, Yuri~S.
  Kivshar, and Boris Luk'yanchuk.
\newblock Optically resonant dielectric nanostructures.
\newblock {\em Science}, 354(6314):aag2472, 2016.

\bibitem{Auguie2008}
Baptiste Augui{\'e} and William~L. Barnes.
\newblock Collective resonances in gold nanoparticle arrays.
\newblock {\em Physical Review Letters}, 101(14):143902, September 2008.

\bibitem{Maier2007}
Stefan~A. Maier.
\newblock {\em Plasmonics: Fundamentals and Applications}.
\newblock Springer, New York, 2007.

\bibitem{Parzefall2015}
Marco Parzefall and Lukas Novotny.
\newblock Optical antennas driven by quantum tunneling.
\newblock {\em ACS Photonics}, 2:1495--1503, 2015.

\bibitem{Svetovoy2014}
V.~B. Svetovoy and G.~Palasantzas.
\newblock Optical rectification in metal-insulator-metal tunnel junctions.
\newblock {\em Physical Review Applied}, 2:024005, 2014.

\bibitem{Nelz2018WaferDiamond}
Richard Nelz, Johannes G\"{o}rlitz, Dennis Herrmann, Abdallah Slablab, Michel
  Challier, Mariusz Radtke, Martin Fischer, Stefan Gsell, Matthias Schreck,
  Christoph Becher, and Elke Neu.
\newblock Toward wafer-scale diamond nano- and quantum technologies.
\newblock {\em APL Materials}, 7(1), January 2019.

\end{thebibliography}
 \end{document}